\documentclass{article}

\usepackage{graphics}
\usepackage{graphicx}
\usepackage{natbib}

\title{Scaling dependence on the fluid viscosity ratio in the selective withdrawal transition}
\author{Itai Cohen}
\date{January 2002}

\begin{document}
\maketitle
\begin{abstract}
In the selective withdrawal experiment fluid is withdrawn through a tube with its tip 
suspended a distance $S$ above a two-fluid interface.  At sufficiently low withdrawal 
rates, $Q$, the interface forms a steady state hump and only the upper fluid is withdrawn.  
When $Q$ is increased (or $S$ decreased), the interface undergoes a transition so that the 
lower fluid is entrained with the upper one, forming a thin steady-state spout.  Near 
this transition the hump curvature becomes very large and displays power-law 
scaling behavior.  This scaling allows for steady-state hump profiles at different flow 
rates and tube heights to be scaled onto a single similarity profile.  I show that the 
scaling behavior is independent of the viscosity ratio.
\end{abstract}

\section{Introduction}
A look at Edgerton's photographic sequences of dripping faucets (see, for example,\cite{Edgerton37}) 
should instantly convince any skeptic that there is something fascinating about a 
fluid interface changing its topology which hints at the richness of the underlying physics.  Much 
attention has been devoted towards trying to classify these topological transitions in fluid 
systems (see, \cite{Calc93,Bertozzi94,Goldstein93,Pugh98}) in the same manner as one 
classifies thermodynamic 
transitions.  It has been shown that when the topological transition involves the formation 
of a singularity  in the fluid flows or interface shapes, as is the case with a drop dripping 
from a faucet, one can use the singularity to organize studies of the 
transition systematically (see, for example, \cite{Barenblatt96,JRL&HAS98,IC&SRN_D01}).  
As the singularity is 
approached, the separation between the length scales characterizing the boundary 
conditions and the length scales characterizing the system grows.  Furthermore, 
the fluid systems can enter an asymptotic regime where a subset of the terms in the governing 
equations dominate the flows.  The organization of the transition 
studies involves the identification of the different subsets of terms which can balance 
as the singularity is approached.  In these asymptotic regimes, the systems typically display 
scaling of length scales or other physical quantities and often, the interfacial profiles near 
the singularity can all be collapsed onto a universal curve.  In many systems these singularities 
manifest themselves in the transition dynamics 
(see, \cite{JRL&HAS98,Bensimon86,Sid&Lene00,Lathrope00}.)  
Here, this approach is extended to the study of the steady-state interface profiles near the 
topological transition associated with the industrially important (see, for example, 
\cite{Muskat49,Bear72,Itai_coating,Ganan-Calvo98}) process of selective withdrawal. 

\begin{figure}[htbp]
\centerline{
\rotatebox{270}{
\resizebox{0.9\textwidth}{!}
{\includegraphics[clip=]{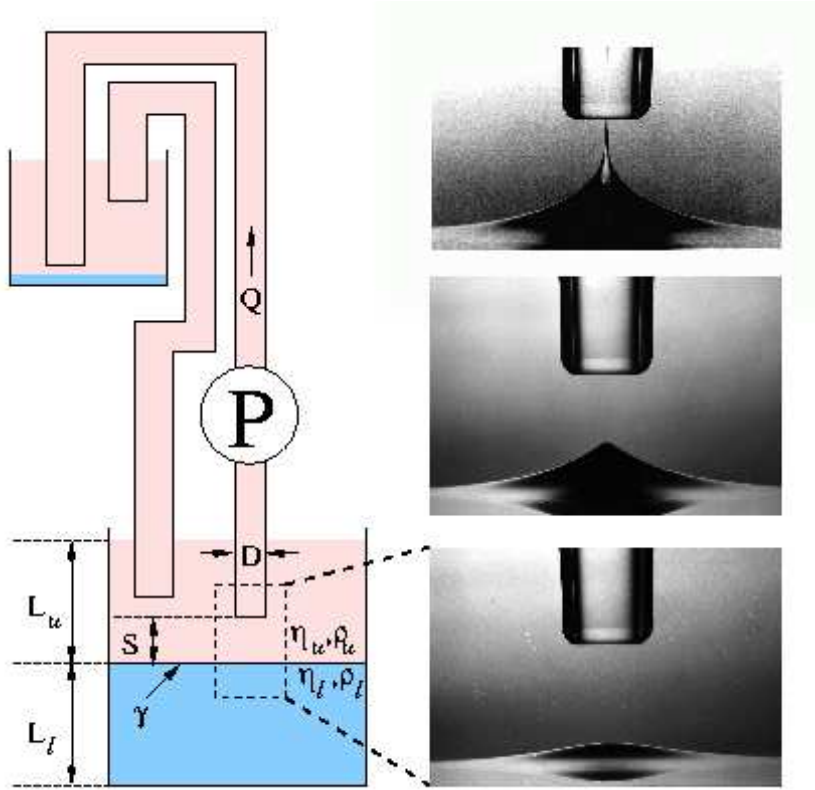}
}}}
\caption
[Diagram of apparatus and photographs of interface at different $Q$]
{Diagram of the experimental apparatus and photographs of the steady state interface at different $Q$.  
Fluid is withdrawn from the withdrawal container and deposited into a waste container.  The upper fluid is 
then siphoned back into the withdrawal container. As shown in the diagram, the parameters for this 
problem include the height of the tube from the interface, $S$, withdrawal rate, $Q$,  surface 
tension, $\gamma$, fluid densities, $\rho_{u}$ and  $\rho_{l}$, fluid viscosities, $\eta_{u}$ and  
$\eta_{l}$, tube diameter, $D$, fluid height above the interface, $L_u$, fluid height 
below the interface, $L_l$, container size, and surfactant concentration. Viewed bottom to top, the photographs show the 
evolution of the steady state interface as $Q$ is increased.  The top photograph of the 
interface in the spout state is taken from \cite{Itai_coating}. The middle photograph of the 
hump at the transition flow rate is taken from \cite{IC&SRN_SW01}.} 
\label{fig:SW_apparatus}
\end{figure}

In the selective withdrawal experiment a tube is immersed in a filled container so that 
its tip is suspended a height $S$ above an interface separating two immiscible fluids (figure \ref{fig:SW_apparatus}).  
When fluid is pumped out through the tube at low flow rates, $Q$, only 
the upper fluid is withdrawn. The flows deform the interface into 
an axi-symmetric steady-state hump with a stagnation point at the hump 
tip (Fig. \ref{fig:SW_apparatus}).  The hump grows in height and curvature as 
$Q$ increases (or $S$ decreases) until the flows undergo a transition where the lower fluid 
becomes entrained in a thin axi-symmetric spout along with the upper fluid.  The interface 
becomes unbounded in the vertical direction, the stagnation point moves from the hump 
tip into the interior of the lower fluid, and the upper fluid geometry becomes toroidal thus 
changing the topology of the steady state.  Once the spout has formed, an increase in $Q$ 
(or decrease in $S$) causes the spout to thicken.

Near the transition, the steady-state mean radius of curvature at the hump tip, $1/\kappa$, can be 
orders of magnitude smaller than the length scales characterizing the boundary conditions 
(for example the tube diameter, $D$).  However, for the range of parameters explored thus far, 
even when the system is arbitrarily close 
to the transition, $\kappa$, while large, remains finite.  The manifestation of a true 
singularity would entail the 
divergence or vanishing of some physical quantities or length scales describing the system.  
In practice, for all physical systems
the approach to the singularity is cut off at some length scale.  For the 
selective withdrawal system where both fluids have a viscosity of about 2 St, \cite{IC&SRN_SW01} 
show that the cutoff appears when the mean radius of curvature at the hump tip is about 200 $\mu$m 
which is well above the breakdown limit for the Navier-Stokes equations.  It is 
still unclear which physical parameters determine this relatively large length scale for the 
hump mean radius of curvature cutoff, $1/\kappa_u$.  

Nevertheless, fixing $S$ and looking at the steady-state profiles as $Q$ is increased, 
Cohen and Nagel observe that, up until the cutoff, both the hump curvature and height 
display scaling behavior characteristic of systems approaching a singularity.  
Furthermore, they are able to use the observed scaling relations to collapse the hump 
profiles near the transition onto a universal curve.  In doing so, they have shown that it is 
possible to treat this as a ``weakly-first-order'' transition and use the 
same systematic approach others have successfully applied to the study of the drop snap-off 
problem\footnote{While scaling has been hypothesized by \cite{Acrivose78} for an analogous 
3-D system it was never shown experimentally.}.  
However, understanding 
how to control the cutoff curvature is crucial for obtaining a more accurate scaling analysis.  
Moreover, a precise understanding of the transition would allow for control of the minimum spout 
diameter.  This control could, in turn be used to advance new emerging technologies such as 
coating micro-particles (\cite{Itai_coating}), creating mono-dispersed micro-spheres 
(\cite{Ganan-Calvo98}), and emulsification through tip streaming (\cite{EggersRev97,Sherwood}) 
which take advantage of the selective withdrawal geometry.

Further insight into this cutoff behavior is gained by comparing with an analogous 
two-dimensional (2-D) problem which roughly corresponds to replacing the tube with a 
line sink. \cite{Moffatt92} showed that in an idealized case where the bottom fluid 
is inviscid while the top fluid is very viscous, the 2-D hump interface forms a 
two-dimensional cusp singularity beyond sufficiently high withdrawal rates.  Recently, \cite{Eggers01}
showed that when the lower fluid has a finite viscosity, the solution to the governing 
equations changes and the system no longer manifests a singularity.  
Instead, the approach to the singularity is 
cut off and the system undergoes a transition to a different steady state.  In this new 
state, a sheet of the lower fluid is entrained along with the upper fluid into the line sink.  
However, the finite lower fluid viscosity prevents the hump profiles from scaling onto a 
similarity solution.  In contrast, \cite{IC&SRN_SW01} point out that even though they 
performed their experiments for 3-D selective withdrawal with an upper and lower fluid 
which were comparably viscous (about 2 St), they still observe scaling behavior and 
profile collapse onto a similarity solution.  Whether or not performing the experiments 
with a lower fluid which is less viscous will allow the system to get closer to the 
singularity (as is the case in the 2-D analogue) has so far remained a mystery.  
In the present paper, a systematic analysis of the role which the lower fluid 
viscosity, $\eta_{l}$,  plays in the selective withdrawal transition shows that 
neither $\eta_{l}$ nor the viscosity ratio affect the scaling dependence of 
the hump profiles and the value of the cut-off curvature $\kappa_u$. 

In addition to studying the various scaling dependencies, \cite{IC&SRN_SW01} map out 
the location of the transition in the $S$ vs. $Q$ parameter space.  Indeed, this topic is by far
the most common feature of the selective withdrawal transition discussed 
in the literature.  Yet, even this seemingly 
benign subject has its subtleties.  For example, the experiments of \cite{Blake&Ivey86}, 
\cite{Jirka&Katavola79}, \cite{Harleman59}, and others 
used two fluids which had different densities but were miscible.  
However, as pointed out by \cite{JRL89}, when surface tension is absent, there is always some 
fraction of the lower fluid which is extracted so that those experiments were not tracking 
the actual withdrawal transition\footnote{While these experiments did not track the transition, they may have been sufficient 
for their intended purpose of modeling magma layer mixing during volcanic eruptions.}.
Cohen and Nagel's studies of a pair of immiscible fluids 
which are comparable in their viscosity (about 2 St) are the 
first to take into account the effects of the surface tension in experiments 
mapping out the transition location for low Reynolds number flow.  
In this paper, these studies are extended to other fluid combinations. 
Once again, I find that neither $\eta_{l}$ nor the viscosity ratio affect
the values of $S$ and $Q$ at which the transition occurs.   

The paper is organized as follows: Section \ref{sec:exp} describes the fluids and experimental 
apparatus used to make the measurements.  This section also describes the effects of the 
container walls and fluid layer thickness on the transition.  Section \ref{sec:trans_struct} 
addresses the hysteretic nature of the transition. Also addresses is the correlation 
between the amount of hysteresis and the final value for the hump 
curvature, $\kappa_{u}$, when the transition takes place at low enough values of $S$. 
In section \ref{sec:scaling} a scaling 
analysis is performed for a system which is identical to that used by Cohen and Nagel 
with the exception that the viscous lower fluid is replaced with one which is two hundred 
times less viscous.  The two sets of results are then compared.  In section \ref{sec:SuvsQ}, 
I map out the 
location of the selective withdrawal transition in the $S$ vs. $Q$ parameter space as a function 
of the upper and lower fluid viscosities.  Also presented is a comparison of the 
experimental results with the theoretical prediction of \cite{JRL89} for the transition 
dependence on the parameters $S$ and $Q$.  Section \ref{sec:conclusion} presents a 
discussion of the results. 
Finally, the role of surfactants is addressed in Appendix A.  Since it is 
extremely difficult to keep a fluid interface free of surfactants it is important to isolate 
and control their effects.  In the current studies the effects on the experimental 
results are found to be rather benign. 

\section{Characterization of Fluids and Experimental Details}\label{sec:exp}

As shown in figure \ref{fig:SW_apparatus}, the parameters important for this experiment are the 
upper and lower fluid viscosities and densities ($\nu_u$, $\nu_l$, $\rho_u$, $\rho_l$), 
the interfacial tension ($\gamma$), 
the orifice diameter ($D$), the height of the orifice ($S$), the flow rate ($Q$), the fluid height 
above the interface ($L_u$), the fluid layer thickness below the interface ($L_l$), the container size, and 
the surfactant concentration.  In order to look at the scaling of the steady-state profiles, 
care must be taken in designing an experimental apparatus capable of isolating the profiles 
near the transition.  Figure \ref{fig:SW_apparatus} displays a diagram of the apparatus in which the 
experiments were performed.  A large tank ($30$ cm $\times$ $30$ cm $\times$ $30$ cm) capable of holding 
fluid layers that were each about 12 cm in height is used as the withdrawal container for 
the fluids.  During the experiments, fluid is pumped out of the withdrawal container and 
into a waste container.  When the system is in the spout state the fluids enter the waste 
container as an emulsion.  This emulsion is deposited at the bottom of the waste container 
which is where the droplets comprised of the lower fluid remain.  The upper fluid is 
siphoned back into the withdrawal container through large tubes at a rate which matches 
the withdrawal rate thereby keeping constant the upper fluid layer thickness in the withdrawal 
container.  Note that the bottom fluid layer does not change its thickness when 
the system is in the hump state and decreases its thickness with time when the system is 
in the spout state.  However, even for thick spouts (0.1 mm) and for large flow rates (10 
ml/sec) the lower fluid layer decreases its thickness at the very slow rate of 0.01 mm/min 
which corresponds to a 0.1$\%$ change in the straw height, $S$.

The fluids were withdrawn using a B­9000 Zenith metering gear pump attached to 
a variable speed DC motor.  A Dynapar Rotopulser encoder was used to read out the 
withdrawal rate.  The pump uses gears to displace fluid from the pump intake to the 
pump outlet.  There are small variations in flow rate associated with the filling and 
draining of the gaps between the gear teeth.  However, at large flow rates, or equivalently 
at high rotation frequencies, these variations damp out and the amplitude of the remaining 
noise corresponds to a very small percentage of the total flow rate.  High rotation rates 
also average out the fluctuations in the driving motor.  Using bigger or smaller pump 
attachments allows for pumping of the fluid at the same flow rate but at a different gear 
rotation rate.  This allows for the determination of the effects of noise in the flow 
rate on the transition.  A further reduction of the noise in the experiments is achieved by 
siphoning the fluids into the waste container.  However, when siphoning, the maximum 
rate of withdrawal (determined by the viscosity, straw diameter, transfer tube length, 
transfer tube diameter, and fluid height difference between the withdrawal and waste 
containers) is substantially smaller. 

The fluids used were heavy mineral oil (HMO), light mineral oil (LMO), silicone oil (polydimethylsiloxane or PDMS), salt water (Salt H$_2$O), and mixtures of glycerin and water (Water/Glycerin).  No surface chemistry was observed at the two-fluid 
interfaces even when the liquids remained in contact for periods longer than a month.  
However, a slight change in the transition flow rate at fixed $S$ over a period of days 
indicated that the surfactant concentration at the interface was increasing with time.  A 
detailed discussion of the surfactant effects appears in Appendix A.  The viscosity, $\nu$, was 
measured using calibrated Cannon Ubbelohde viscometers immersed in a Cannon constant 
temperature bath.  In this manner the viscosity could be determined to within $\pm 5 \%$.  
Glycerin can be diluted with water so that the resultant fluid has $0.01 \leq \nu \leq 9.8$ St (\cite{Segur51}).  
Table \ref{table:System_properties} lists the values of the viscosities and densities for the different fluids.

The surface tension, $\gamma$, of the two-fluid interface was determined using the 
pendant drop method (see, for example, \cite{Neumann95,Hansen91}) which takes advantage of the competition between the 
surface tension and buoyant forces acting on a static drop hanging from a nozzle.  The 
buoyant forces distort the drop from a spherical shape.  Measuring the distortion and 
density mismatch allows a determination of the surface tension.  Implementation of 
this technique on water, toluene, and di-methylformamide showed a capability for
measuring the surface tensions to within $\pm 10 \%$.   Table \ref{table:System_properties} lists 
the values of $\gamma$ for the different fluids.

\begin{table}[htbp]
\centerline{
\resizebox{1.0\textwidth}{!}
{\includegraphics{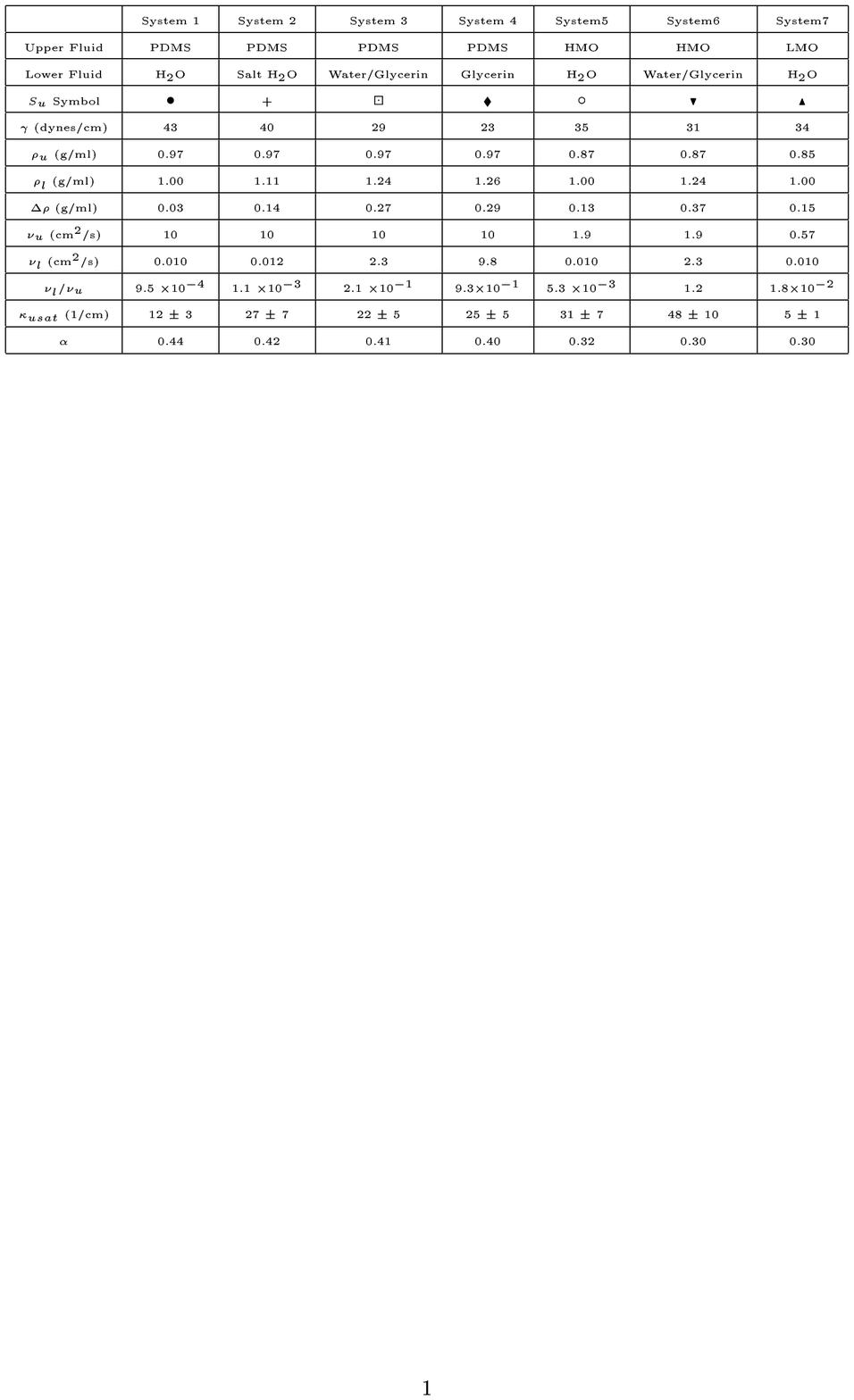}}}
\caption[List of the properties for fluid systems studied]
{List of the properties for each fluid system studied. Row 1 and 2 list the fluids used. Row 3 lists 
the symbols used in plotting the experimentally measured $S_u$ curves in figures \ref{fig:Su_vs_Q_compare}, 
 \ref{fig:Su_vs_Q_theory_compare}, \ref{fig:Su,Ku,hu,DeltaSvsQ}, and \ref{fig:Diameter_dependence}. 
Rows 4 through 10 list the values of the fluid parameters 
measured for the different systems. The second to last row shows the asymptotic 
value at which the mean curvature saturates, $\kappa_{usat}$, and the last row 
lists the power-law exponent $\alpha$ used in fitting the experimentally measured $S_u$
curves in figures \ref{fig:Su_vs_Q_compare}, \ref{fig:Su_vs_Q_theory_compare}, 
\ref{fig:Su,Ku,hu,DeltaSvsQ}, and \ref{fig:Diameter_dependence}
}
\label{table:System_properties}
\end{table}

The apparatus was illuminated from the rear and a CCD video camera was used to image 
silhouettes of the steady state hump shapes.  The images were transferred onto a PC 
where an edge tracing IDL program tracked and recorded the points where the derivative 
of the pixel intensity profile across the hump interface was extremized.  The profiles were 
then superimposed onto the original images and checked for accuracy.  In order to 
determine the mean curvature at the hump tip, $\kappa$, a Gaussian function was 
used to fit the tip of the recorded hump profile. The value of $\kappa$ is taken to be 
the curvature of the fitting function at the hump tip. 

\begin{figure}[htbp]
\centerline{
\resizebox{0.5\textwidth}{!}
{\includegraphics{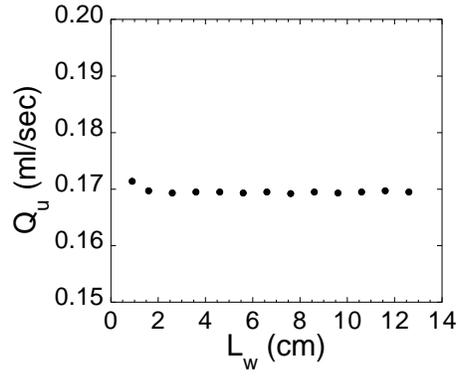}}}
\caption[Plot of $Q_u$ versus $L_w$]
{Plot of the transition flow rate, $Q_u$ as a function of the distance 
between the withdrawal tube and the container wall, $L_w$ for system 3 in Table 
\ref{table:System_properties}.  Beyond 2 cm, $Q_u$ shows no dependence on $L_w$. 
Note that the Value of $Q_u$ is plotted on a linear scale.
}
\label{fig:Qu_vs_Lw}
\end{figure}

\begin{figure}[htbp]
\centerline{
\resizebox{0.5\textwidth}{!}
{\includegraphics[clip=]{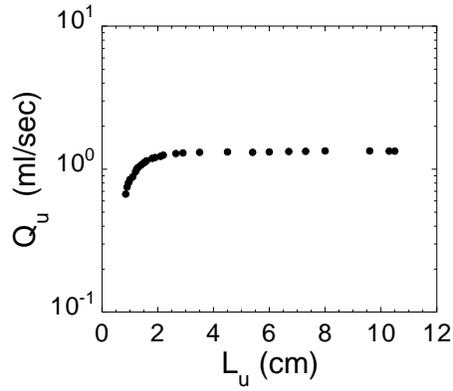}}}
\caption[Plot of $Q_u$ versus $L_u$]
{Plot of the transition flow rate, $Q_u$ as a function of the upper fluid layer 
thickness, $L_u$ for system 5 in Table \ref{table:System_properties}.  Beyond 3 cm, $Q_u$ shows no dependence on $L_u$.
}
\label{fig:Qu_vs_Lu}
\end{figure}

Using this apparatus, I tested the effects of the container walls and thickness of 
the upper fluid layer on the transition flow rate, $Q_u$.  Figure \ref{fig:Qu_vs_Lw} 
shows a plot of the 
transition flow rate $Q_u$ vs. the distance $L_w$ from one of the container walls while the straw 
height, $S$, is held constant.  The fluid parameters for this particular experiment correspond to 
those of system 3 in Table \ref{table:System_properties}. When the straw is farther than 
about 2 cm from the 
container wall, there is no variation in the transition flow rate.  Figure \ref{fig:Qu_vs_Lu} shows a plot 
of the transition flow rate $Q_u$ as a function of the upper fluid layer thickness $L_u$ for 
constant $S$.  The fluid parameters for this particular experiment correspond 
to those of system 5 in Table  \ref{table:System_properties}.  
As $L_u$ is increased, $Q_u$ first increases but eventually saturates and remains 
constant for $L_u$ greater than about 3cm. In all of the experiments, the bottom 
fluid layer thickness $L_l$ was typically kept at about 12 cm.  Measurements of 
the transition flow rates showed no significant variations between systems with 
an $L_l$ of 12, 10, and 6 cm. Similar measurements for the $Q_u$ dependence on 
$L_w$, $L_u$, and $L_l$ were conducted for all of the different 
fluid combinations used in the experiments.  For the tube diameters ($D = 0.16$ cm and $D = 
0.79$ cm), tube heights ($0.07$ cm $< S < 2.0$ cm), and flow rates ($Q < 20$ 
ml/sec) used in the experiments, the container walls were always sufficiently distant and 
the fluid layers were always sufficiently thick so as not to affect the flows.

\section{Transition Structure and Hysteresis}\label{sec:trans_struct}

In their paper, \cite{IC&SRN_SW01} map out the location and structure of 
the selective withdrawal transition for a pair of fluids corresponding to system 6 in 
table \ref{table:System_properties}. Here I review these results and note that 
all of the fluid systems in table \ref{table:System_properties} display the same qualitative 
features.  Much of our understanding of the selective withdrawal transition can 
be conveyed by focusing on the effects of the parameters $S$ and $Q$. By fixing $Q$ and tracking  
the hump profiles the tube height below which the interface forms a spout, $S_{u}$, 
can be determined. Figure \ref{fig:Su,Ku,hu,DeltaSvsQ}(taken from \cite{IC&SRN_SW01}) 
shows that $S_{u} \propto Q^{0.30 \pm 0.05}$
(A detailed discussion of how this result changes with the various fluid parameters 
along with a comparison to currently available theory is presented in section \ref{sec:SuvsQ}).
As mentioned in the introduction, for the entire range of parameters 
explored thus far, the evolution of the steady 
state hump profiles is cut off by the hump to spout transition before $\kappa$ diverges.
However it is possible that there is more than one mechanism that is responsible for the transition 
cutoff. If this is the case it is important to isolate the regimes over which particular 
mechanisms dominate.  The cut-off suggests that the transition may be hysteretic.
Indeed, for transitions occurring at low $Q$, or equivalently at small $S_u$, 
the straw height at which the spout decays back into a hump is different from $S_u$.
The difference of the two heights or hysteresis is defined as $\Delta S$.  
Figure \ref{fig:Su,Ku,hu,DeltaSvsQ} shows that $\Delta S$ decreases exponentially with 
the flow rate $Q$.  For this 
particular system the hysteresis can be fit with the function: $\Delta S  = 0.04 \exp^{-Q/0.032}$.  
For $Q > 0.1$ ml/sec, $\Delta S$ was too small to measure.  Figure \ref{fig:Su,Ku,hu,DeltaSvsQ} also shows how the hump 
mean radius of curvature, $1/\kappa_u$, and height, $h_u$, at the transition vary with $Q$.  The dramatic 
decrease in $\Delta S$ coincides with the onset of a flat asymptotic dependence for $1/\kappa_u$ at $Q > 
0.1$ ml/sec.  In order to quantify this correlation, the curvature data is fit with the form 
$1/\kappa_u = 0.02 + 0.32\exp^{-Q/0.032}$ which has the same exponential decay with $Q$ as does the 
hysteresis. The curvature saturation values, $\kappa_{usat}$, for all of the systems are 
shown in the second to last row in Table \ref{table:System_properties}.

\begin{figure}[htbp]
\centerline{
\resizebox{0.9\textwidth}{!}
{\includegraphics[clip=]{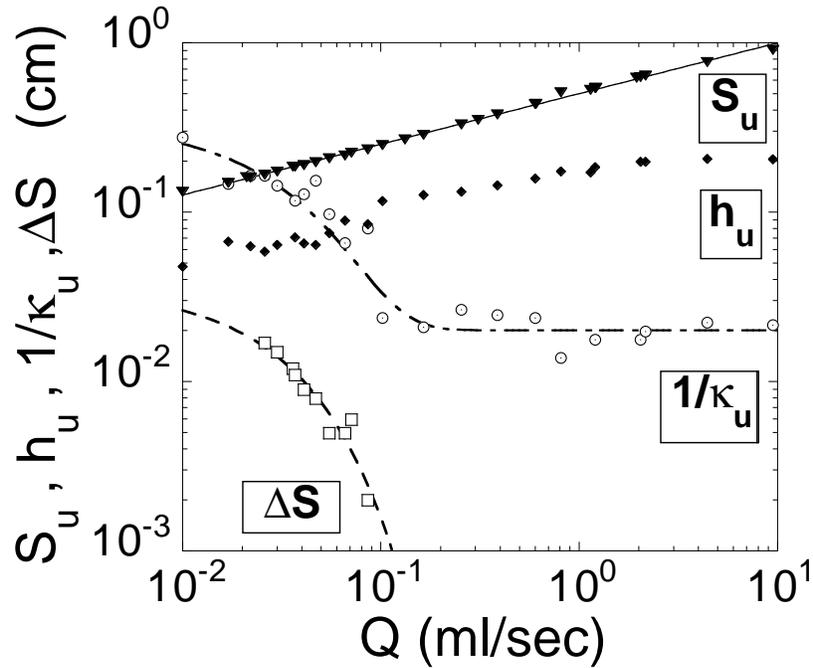}}}
\caption[Plots of $S_u$, $\Delta S$, $h_{u}$, and $1/\kappa_{u}$ as a function of $Q$]
{
Plots of the transition tube height, $S_{u}$, hysteresis, $\Delta S$, transition hump height, $h_{u}$, and transition radius 
of curvature $1/\kappa_{u}$, as a function of the flow rate $Q$. $S_{u} \propto Q^{0.3 \pm 0.05}$ 
(solid line).  The $\Delta S$ data is fit with an exponential decay (dashed) of the form: $\Delta S = 
0.04 \exp^{-Q/0.032}$ although the range is not sufficient to exclude a power-law decay.  
The mean radius of curvature, $1/\kappa_{u}$ is fit (dash dot) with the form: $1/\kappa_{u} = 0.02 + 0.32 \exp^{-Q/0.032}$.
This figure is reproduced from \cite{IC&SRN_SW01}. 
}
\label{fig:Su,Ku,hu,DeltaSvsQ}
\end{figure}

\begin{figure}[htbp]
\centerline{
\resizebox{0.7\textwidth}{!}
{\includegraphics[clip=]{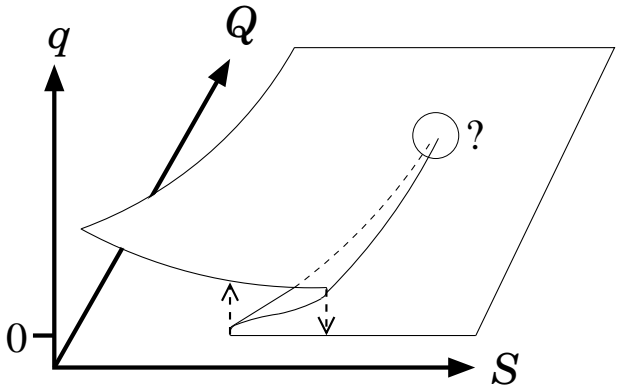}}}
\caption[Bifurcation diagram summarizing the features of the selective withdrawal transition]
{Bifurcation diagram summarizing the features of the selective withdrawal transition. 
The vertical axis labeled $q$ depicts the amount of the lower fluid withdrawn as a function of 
the withdrawal rate $Q$ and straw height $S$.  The folded two dimensional sheet represents the 
states which are accessible to the system. The fold represents the amount of hysteresis in the 
transition which becomes too small to measure at large enough $Q$ and $S$.    
}
\label{fig:Transition_structure}
\end{figure}

A diagram (\cite{Hale&Kocak91}) summarizing the features of the transition is shown in figure 
\ref{fig:Transition_structure}.  The vertical axis labeled $q$ 
relates how much of the lower fluid is withdrawn as a function of the total fluid 
withdrawal rate $Q$ and straw height $S$.  The states accessible to the system are represented 
by a sheet which is embedded in this three dimensional parameter space.  The part of the 
sheet which corresponds to the system in the hump state is located in the $q = 0$ plane.  
An experiment where the flow rate is held fixed and the straw height is slowly reduced 
would correspond to a path which traverses the sheet in a direction which is parallel to 
the $S$ axis.  The fold in the sheet represents the hysteresis in the transition.  As the straw 
height is reduced, the system runs out of hump states and makes a discontinuous jump at 
$S = S_u$ (depicted by the dashed arrow pointing upward) to a part of the sheet which has a 
finite value of $q$.  If $S$ is reduced further, the sheet slopes towards higher values of $q$ and 
more of the lower fluid is withdrawn.  If on the other hand, $S$ is increased, the system 
remains on the upper branch of the sheet until the spout states run out at which point the 
system makes another discontinuous jump (depicted by the downward dashed arrow) to 
a part of the sheet where $q = 0$.  The separation between the two arrows along the $S$ axis 
corresponds to $\Delta S$.  As the diagram indicates, at large values of $Q$ and $S$, the amplitude of 
the fold or equivalently $\Delta S$ decreases dramatically.

It is unclear whether at high enough $Q$ or $S$ the system undergoes a bifurcation 
where the hysteresis vanishes, or if there always remains some small fold in the sheet 
which is too small to be detected by the experiments.  It seems unlikely that the transition
could have a cut-off (and remain discontinuous) and yet have no hysteresis.  Such a situation 
would entail the 
disappearance of the hump solutions at the exact location in the parameter space where the 
spout solutions emerge.  It is likely that numerical investigations which are able to detect 
both stable and unstable solutions will be needed in order to answer this 
question definitively.  The exponential decay in both the $\Delta S$ and $1/\kappa_u$ 
curves suggests that even if 
there does exists a small amount of hysteresis at large $S$ or $Q$, there is a separate and 
localized mechanism that generates the transition hysteresis at low values of $S$ or $Q$.  

\begin{figure}[htbp]
\centerline{
\resizebox{0.9\textwidth}{!}
{\includegraphics[clip=]{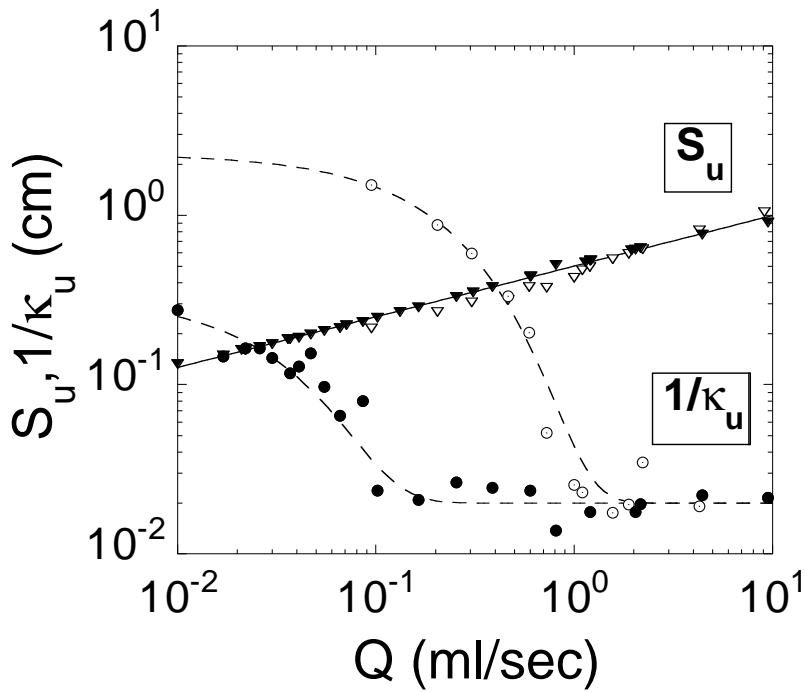}}}
\caption[Dependence of the $S_u$ and $1/\kappa$ vs. $Q$ curves on D]
{ 
Dependence of the $S_u$ and $1/\kappa$ vs. $Q$ curves on the withdrawal tube diameter, D.  
The closed symbols reproduce the $S_u$ and $\kappa_u$ curves in figure \ref{fig:Su,Ku,hu,DeltaSvsQ} 
where $D = 0.16$ cm.  The open symbols depict the values of $S_u$ and $\kappa_u$ 
as a function of $Q$ when $D = 0.79$ cm. The solid line is a fit to the $D=0.16$ 
data which shows that $S_{u} \propto Q^{0.3 \pm 0.05}$. The mean radius of 
curvatures, $1/\kappa_{u}$ are fit (dash) with the forms: $1/\kappa_{u} = 0.02 + 0.32 \exp^{-Q/0.032}$ 
and $1/\kappa_{u} = 0.02 + 2.3 \exp^{-Q/0.22}$ for the $D=0.16$ and $D=0.79$ cm data sets respectively.
}
\label{fig:Diameter_dependence}
\end{figure}

When transitions occur at low $S$ or $Q$, the ratio of the withdrawal tube diameter, $D$, 
over the straw height, $S$, is of order one.  As a result, the circulation rolls set up by 
withdrawal press against the two-fluid interface and the streamlines near the hump tip 
become distorted.  By changing the flow geometry, the straw diameter sets a length scale 
for the hump mean radius of curvature at the transition, $1/\kappa_u$.  
Figure \ref{fig:Diameter_dependence} shows 
the effect of a factor of five increase in $D$ on the $S_u$ and $1/\kappa_u$ curves.  For the range of 
flow rates explored, $\Delta S$ is always about an order of magnitude smaller than $S_u$.  
Therefore, it is not surprising that the two $S_u$ curves in figure \ref{fig:Diameter_dependence} show no significant 
dependence on $D$.  On the other hand, the $1/\kappa_u$ data show that the 
onset of the flat asymptotic dependence for $1/\kappa_u$ occurs at higher $Q$ for larger straw 
diameters.  This evidence indicates that the hysteresis observed in the experiments results 
from the finite width of the withdrawal tube.  At large values of $S_u$ these finite size 
effects vanish. We therefore restrict our scaling analysis to this regime.  

\section{Scaling Analysis And Similarity Solutions}\label{sec:scaling}
Although the cut-off in the transition exists even for high values of $Q$, \cite{IC&SRN_SW01} 
showed that the separation in 
length scales between the hump radius of curvature and the boundary conditions is 
sufficient for the system to display scaling and a similarity solution near the transition.  
As mentioned in the introduction, in the 2-D analogue to the selective 
withdrawal problem the viscosity of the lower fluid, $\nu_l$, greatly affected the observed scaling 
dependencies. In order to determine the effects of $\nu_l$ on the transition structure 
in the present 3-D problem, a comparison of the scaling behavior for two systems with 
different values of $\nu_l$ is presented below. 

\begin{figure}[htbp]

\centerline{
\resizebox{0.5\textwidth}{!}
{\includegraphics[clip=]{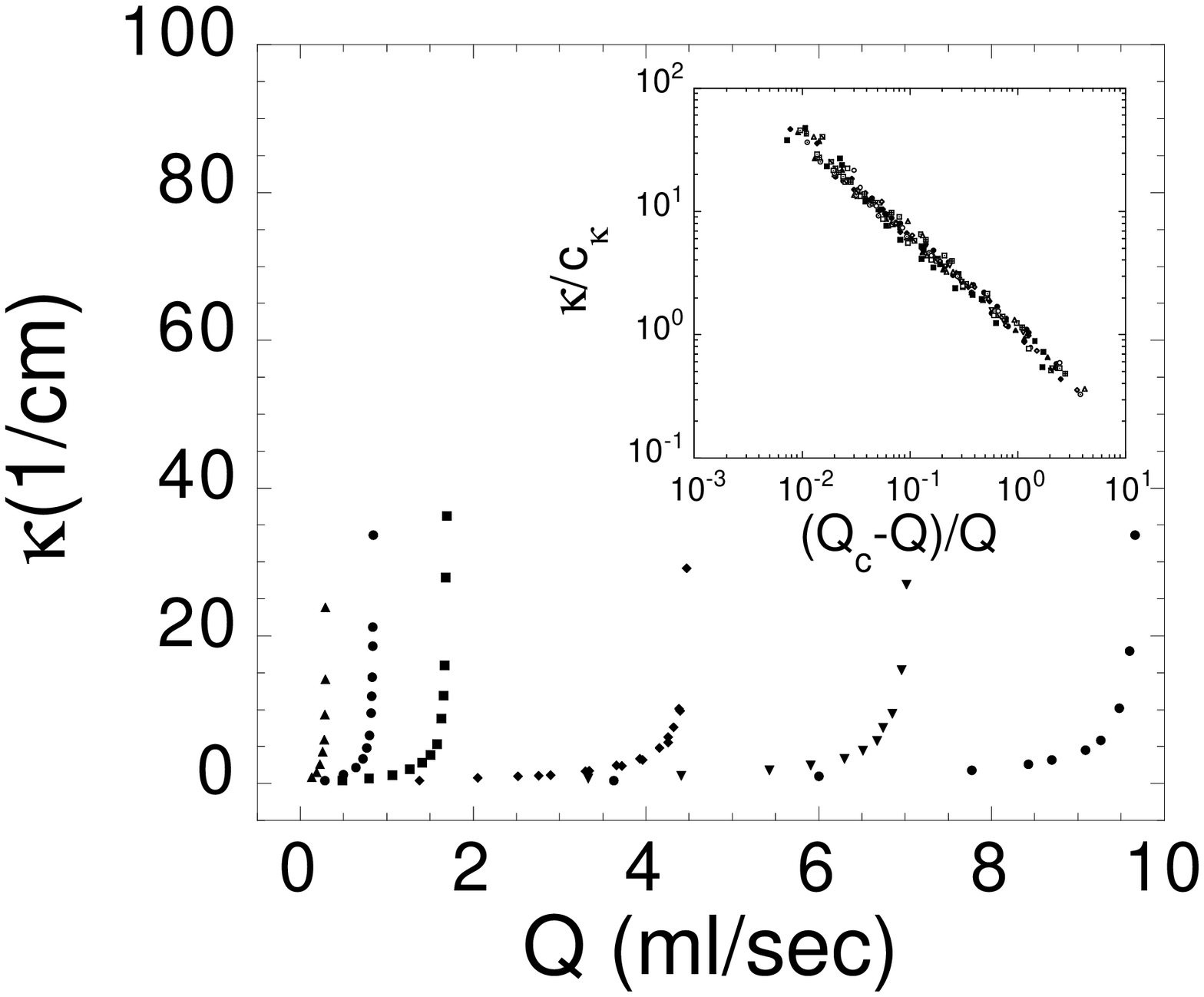}}
\hspace{4mm}
\resizebox{0.5\textwidth}{!}
{\includegraphics[clip=]{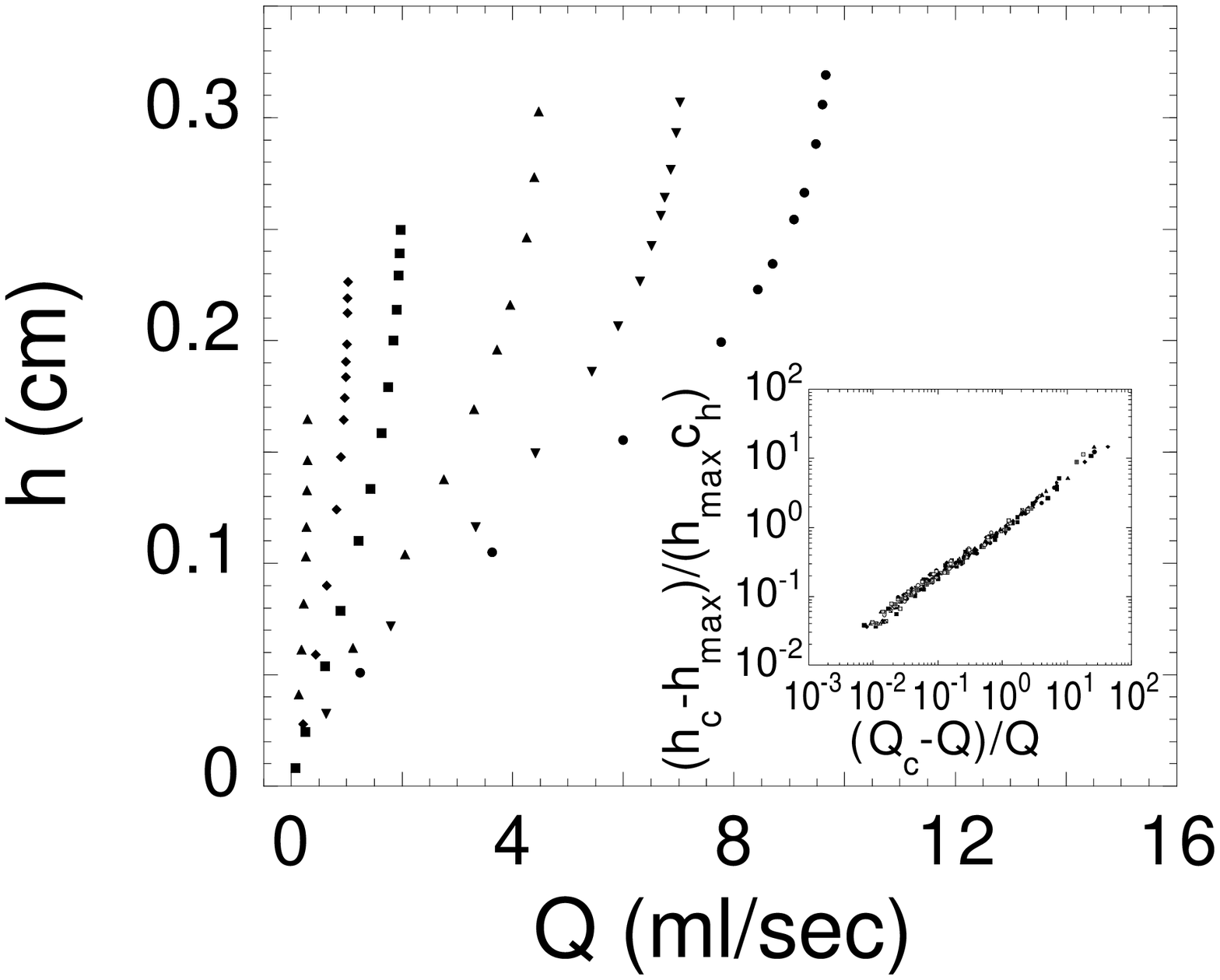}}}
\vspace{4mm}
\centerline{
\resizebox{0.5\textwidth}{!}
{\includegraphics[clip=]{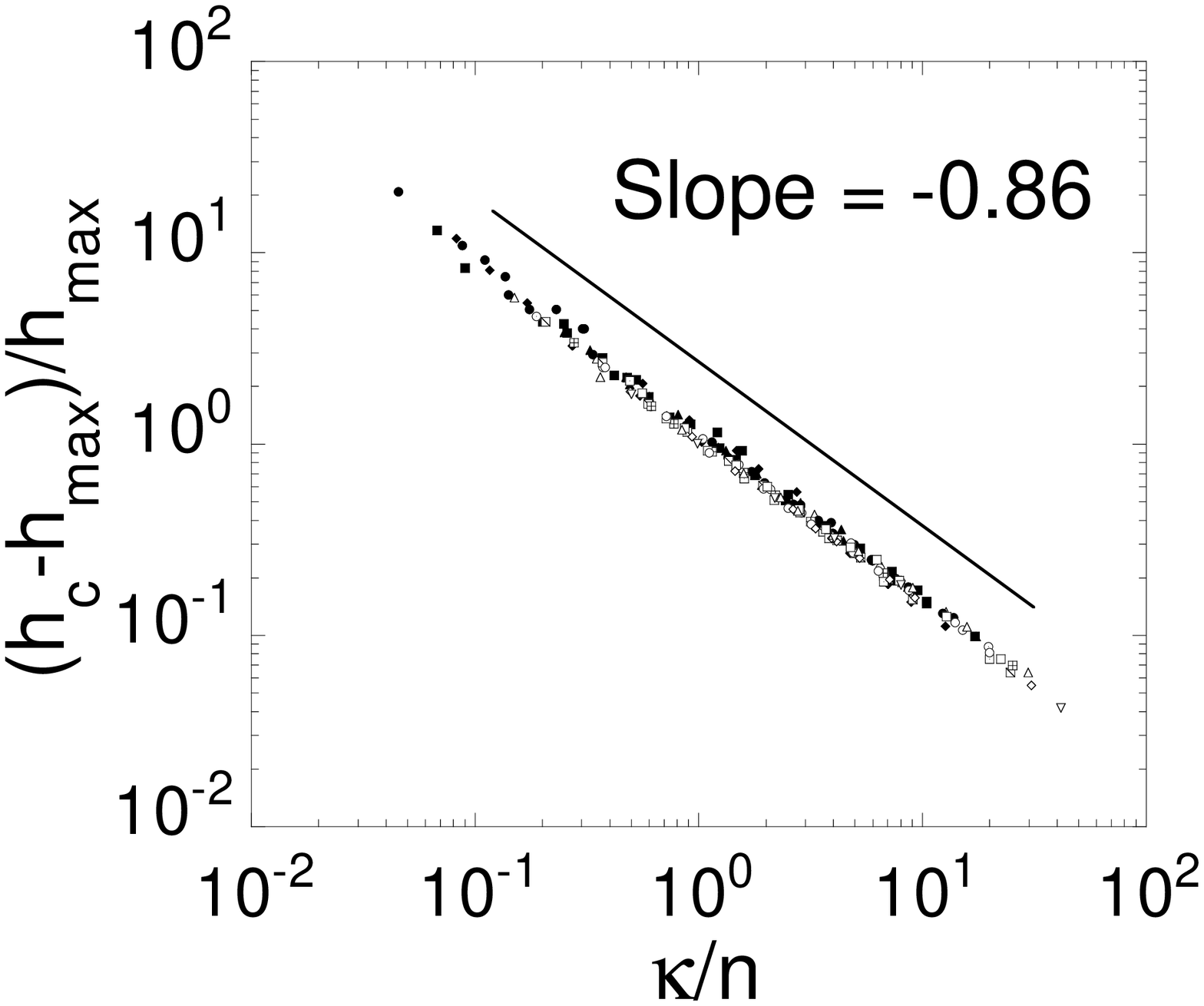}}}
\caption[Scaling of hump mean curvature, $\kappa$, and height $h_{max}$ for system 5]{
Scaling for the hump mean curvature, $\kappa$, and height $h_{max}$. Figure\ref{fig:Scaling200/1}a 
plots the $\kappa$ vs. $Q$ for six tube heights.  Each curve displays 
diverging behavior with increasing $Q$.  For each tube height, a critical flow rate, 
$Q_{c}$, is chosen as a fitting parameter so that as $Q$ approaches $Q_{c}$, the curvatures 
increase as power laws with an exponent of about -0.80 (\ref{fig:Scaling200/1}a inset).  
The prefactors to the curvature power laws, $c_{\kappa}(S)$, are scaled out  
in the inset.  Figure \ref{fig:Scaling200/1}b plots $h_{max}$ vs. $Q$ for six tube heights.  For each $S$, 
a critical hump height, $h_{c}$, is chosen as a fitting parameter. As the flow rate 
approaches $Q_{c}$ (obtained from \ref{fig:Scaling200/1}a inset), the hump heights approach the critical heights as 
power-laws with an exponent of about 0.70.  The prefactors, $c_{h}(S)$, to the power laws in the inset are scaled out.  Figure 
\ref{fig:Scaling200/1}c plots $(h_{c}-h_{max})/h_{max}$ vs. $\kappa/n$ for the entire data set corresponding to twenty 
different straw heights.  The prefactor $n$ 
roughly decreases as $\exp^{-2.5 S}$.  The line corresponds to a 
power-law with an exponent of -0.86.
}
\label{fig:Scaling200/1}
\end{figure}

Figure \ref{fig:Scaling200/200} reproduces the Cohen and Nagel scaling curves for the 
hump curvature and height for system 6 while figure \ref{fig:Scaling200/1} shows the 
results of an identical treatment performed on data taken for system 5 
(where $\nu_l$ is two hundred times smaller).  Figures \ref{fig:Scaling200/200}a 
and \ref{fig:Scaling200/1}a,  show a plot of the mean curvature at the hump tip, $\kappa$ 
(where $0 \leq \kappa \leq \kappa_u$), as a function of $Q$ for 
six representative data sets each of which corresponds to measurements taken at a different 
fixed straw height $S$.  All of the curves display a steep rise in the curvature with increasing 
$Q$.  The insets to figures  \ref{fig:Scaling200/200}a and  \ref{fig:Scaling200/1}a show 
that, as the flow rate 
approaches $Q_c$ (used as a fitting parameter), the steep rise in the curvatures takes the form of 
power-law divergences with exponents of about -0.84 and -0.80 respectively.  For each fluid system, 
all of the data sets (15 sets for large $\nu_{l}$ 
system and 20 for low $\nu_{l}$ system) have the same power-law exponent for the 
divergence.  However, the power-law prefactors, $c_{\kappa}(S)$, change with $S$.  These prefactors 
are scaled out in the insets for clarity.

Figures \ref{fig:Scaling200/200}b and \ref{fig:Scaling200/1}b, show a plot of the hump 
height, $h_{max}$ (where $0 \leq 
h_{max} \leq h_u$), as a function of $Q$ for six data sets each of which corresponds to 
measurements taken at a different fixed straw height $S$.  The insets to figures 
\ref{fig:Scaling200/200}b and \ref{fig:Scaling200/1}b show that as the flow rate 
approaches $Q_c$ (obtained from insets to figures \ref{fig:Scaling200/200}a 
and \ref{fig:Scaling200/1}a), 
the hump heights approache the height $h_c$ (used as a fitting parameter) as power-laws 
with exponents of about 0.73 and 0.70 respectively.  While the 
curves for each fluid system show that $h_{max}$ approaches $h_c$ with the same power-law 
exponent, the power-law prefactors $c_{h}(S)$ change with $S$.  These prefactors are scaled out 
in the insets for clarity.  

We combine the two scaling dependencies in figures \ref{fig:Scaling200/200}c 
and \ref{fig:Scaling200/1}c and plot $\frac{h_c-h_{max}}{h_{max}}$ as a function of 
the normalized curvature $\frac{\kappa}{n}$.  The normalization constant, $n$, 
scales out the power-law prefactors for curves corresponding to different straw heights.  
The two figures show that $\frac{h_c-h_{max}}{h_{max}}$ scales 
as $(\frac{\kappa}{n})^\beta$ where $\beta = -0.85 \pm 
0.09$ for the high $\nu_l$ system and $\beta = -0.86 \pm 0.10$ for the low $\nu_l$ system. 
These results indicate that for this range of straw heights, both 
fluid systems show no change in the power-law exponent $-\beta$ even though the height $h_c$ and 
the power-law prefactor $n$ change with $S$.

\begin{figure}[htbp]
\centerline{\resizebox{0.5\textwidth}{!}
{\includegraphics[clip=]{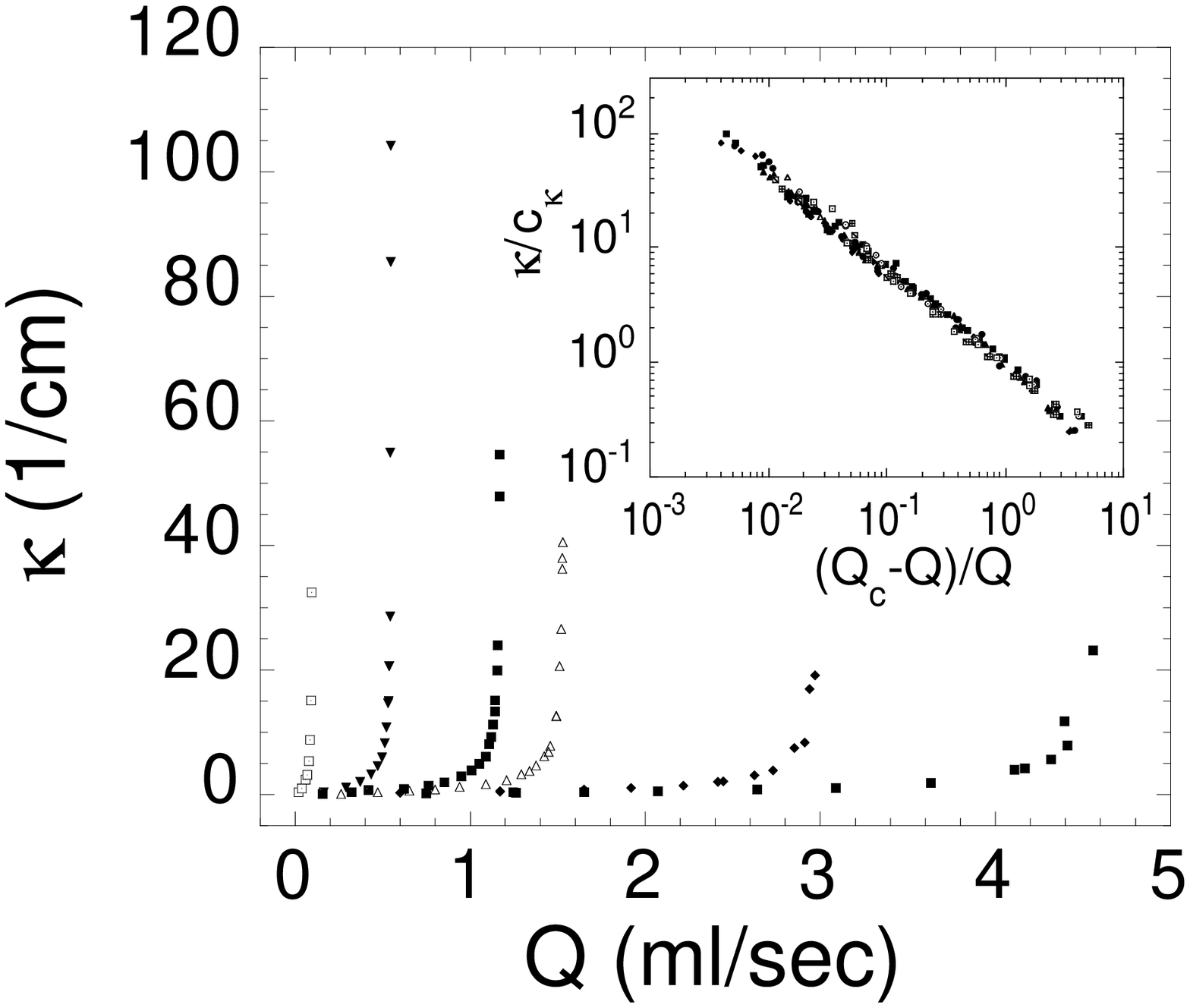}}
\hspace{4mm}
\resizebox{0.5\textwidth}{!}
{\includegraphics[clip=]{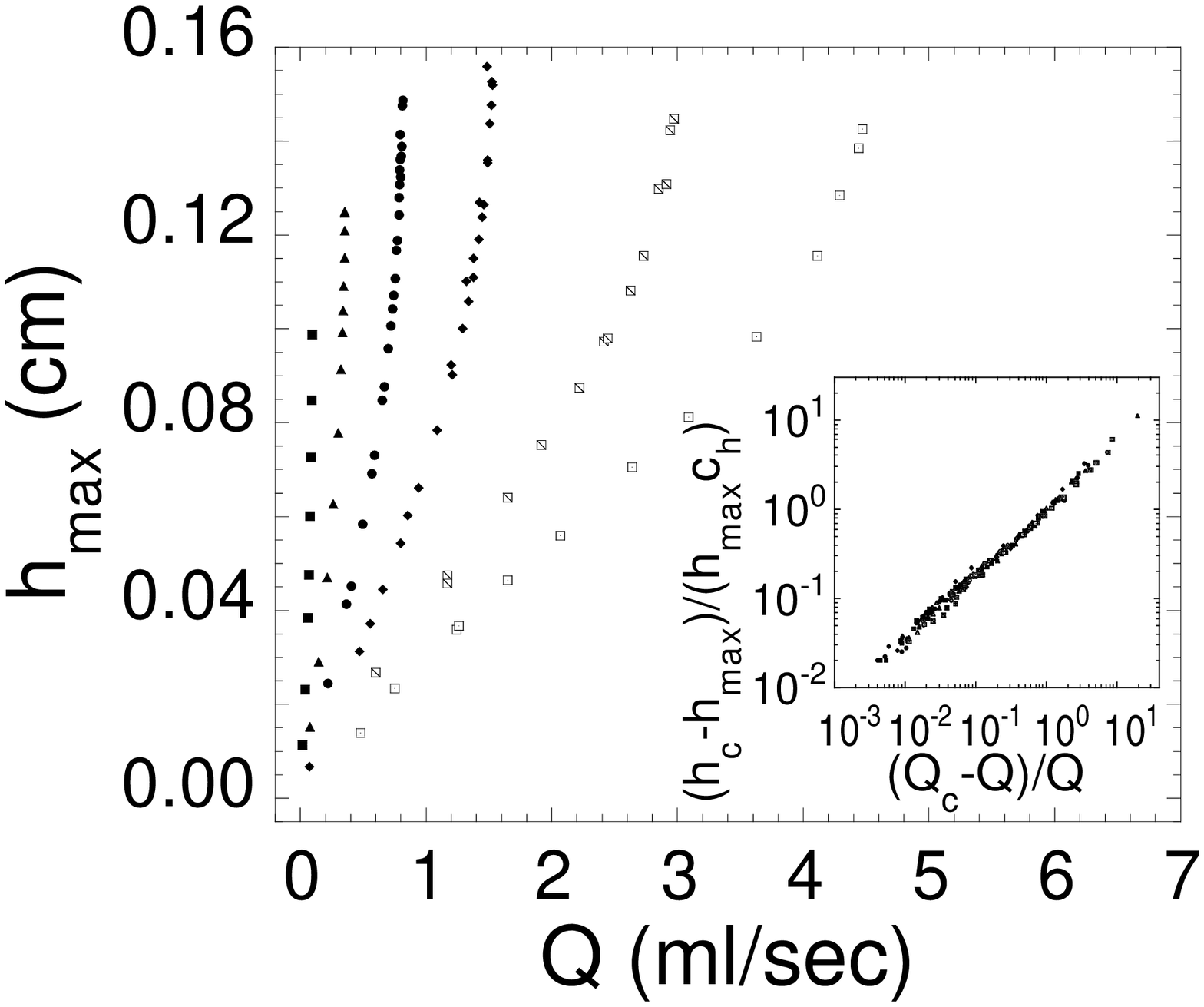}}}
\vspace{4mm}

\centerline{
\resizebox{0.5\textwidth}{!}
{\includegraphics[clip=]{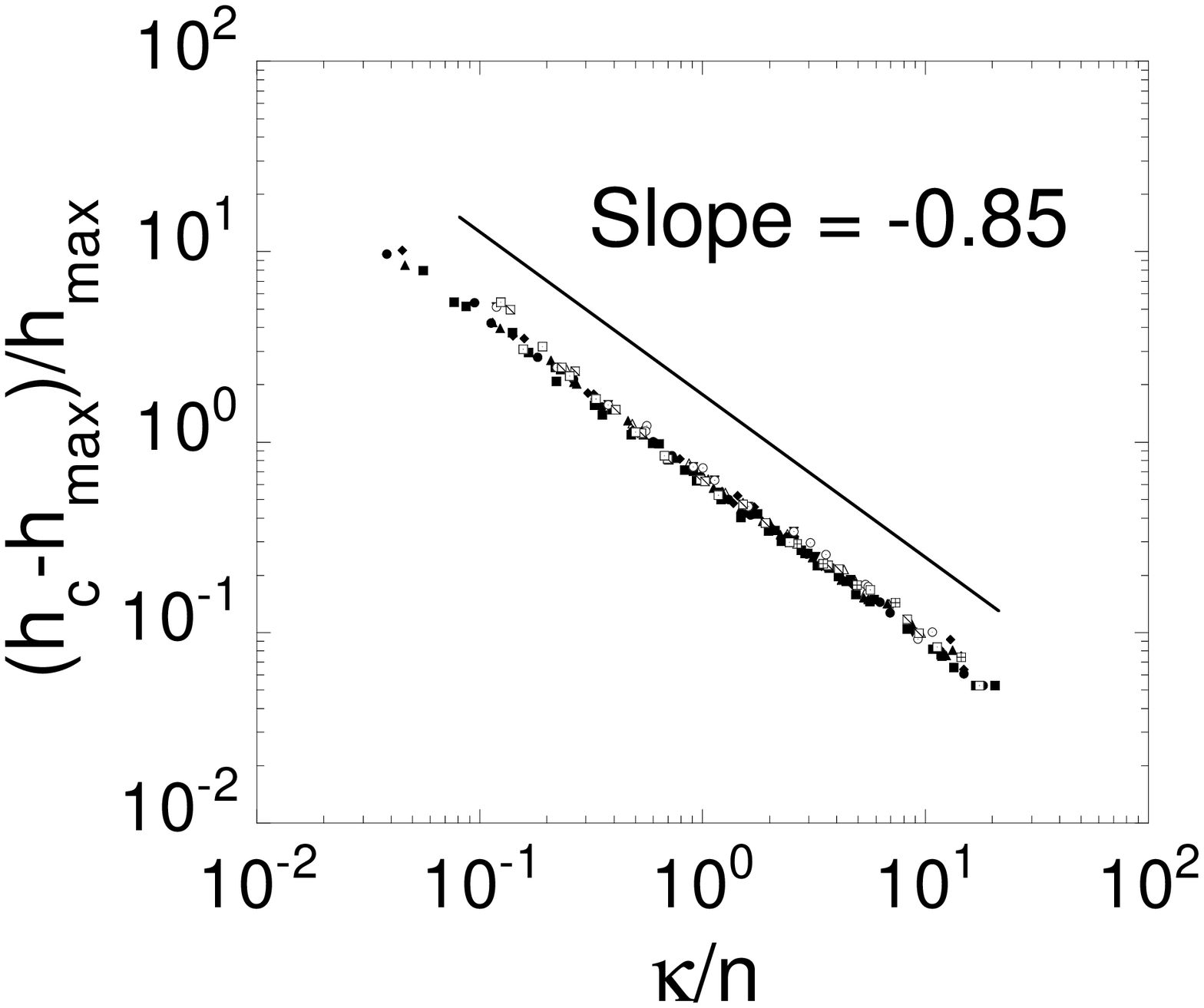}}}

\caption[Scaling of hump mean curvature, $\kappa$, and height $h_{max}$ for system 6]{
Scaling for the hump mean curvature, $\kappa$, and height $h_{max}$.  Figure \ref{fig:Scaling200/200}a 
plots $\kappa$ vs. $Q$ for six tube heights.  Each curve displays 
diverging behavior with increasing $Q$.  For each tube height, a critical flow rate, 
$Q_{c}$, is chosen as a fitting parameter so that as $Q$ approaches $Q_{c}$, the curvatures 
increase as power laws with an exponent of about -0.84 
(\ref{fig:Scaling200/200}a inset).  The prefactors to the curvature 
power laws, $c_{\kappa}(S)$, are scaled out in the inset. 
Figure \ref{fig:Scaling200/200}b plots $h_{max}$ vs. $Q$.  For each $S$, 
a critical hump height, $h_{c}$, is chosen as a fitting parameter. As the flow rate 
approaches $Q_{c}$ (obtained from \ref{fig:Scaling200/200}a inset), the hump heights approach the critical heights as 
power-laws with an exponent about 0.73.  The prefactors, $c_{h}(S)$, to the power laws in the inset are scaled out.  Figure 
\ref{fig:Scaling200/200}c plots $(h_{c}-h_{max})/h_{max}$ vs. $\kappa/n$ for the entire data set corresponding to fifteen 
different straw heights.  The prefactor $n$ 
roughly decreases as $\exp^{-2.5 S}$.  The line corresponds to a 
power-law with an exponent of -0.85. This figure is reproduced from \cite{IC&SRN_SW01}.
}
\label{fig:Scaling200/200}
\end{figure}

Note that $n(S) = c_{h}(S)[c_{k}(S)^\beta]$. The constants $n(S)$ decreases roughly as $\exp^{-
2.5S}$ for both systems. 
While both systems have the same form for the decay of the constants $n(S)$, in the 
lower $\nu_l$ system, the values of $n(S)$ are larger by about ten percent.  Also, as indicated by
table \ref{table:System_properties} , the asymptotic value of $k_u$ at the high $Q$ regime is different by 
about a factor of 1.5.  A detailed analysis of table \ref{table:System_properties} described in Section 6 suggests 
that these very slight differences are attributable the different values 
$\Delta \rho$ in the two systems.  This comparison demonstrates that the lower fluid 
viscosity $\nu_l$ does not affect the scaling relations near the selective withdrawal transition.

The scaling analysis shows that the hump profiles behave as though they are 
approaching a singular solution where, at the flow rate $Q_c$ the hump height would be equal to $h_c$ 
and the mean curvature, $\kappa$, would diverge.  The transition cuts off the evolution of the hump states 
preventing the system from getting arbitrarily close to the singularity and limiting the 
precision with which the power-law exponents can be determined.  $Q_c$ changes with $S$ 
indicating that the system can approach a continuous line of singularities which closely 
tracks the hump to spout transition curve in the $S$ vs. $Q$ parameter space.  

\cite{IC&SRN_SW01} showed that the scaling relations for the hump height and 
curvature can be used to collapse the 
hump profiles near the transition onto a universal curve .  The quantities $\frac{n}{\kappa}$ 
and $\frac{h_c-h_{max}}{h_{max}}$ track how quickly the radial and axial length scales 
decrease as the system approaches the singularity.  In accordance with the definitions of 
Cohen and Nagel, the scaled variables are defined as:

\begin{equation}
\label{similarity_var}
H(R) = \frac{h_{c}-h(r)}{h_{c}-h_{max}} \qquad and \qquad R = \frac{r\kappa}{n},
\end{equation}

\begin{figure}[htbp]
\centerline{
\resizebox{0.9\textwidth}{!}
{\includegraphics[clip=]{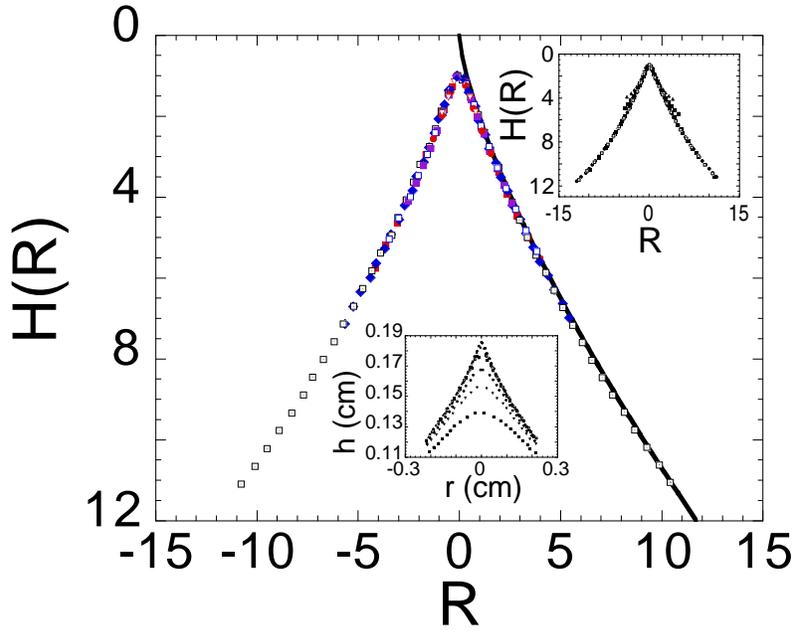}}}
\caption[The scaled hump profiles for system 6]{
The scaled hump profiles for system 6.  The lower inset shows eight profiles taken 
from the $S = 0.830$ data set.  The main figure shows the same profiles after scaling.  The 
solid line corresponds to a power law of the form $R^{0.72}$.  In the upper inset we compare 
the universal curves for the $S = 0.830$ cm, $0.613$ cm, $0.508$ cm, $0.381$ cm, $0.255$ cm data 
sets. This figure is reproduced from \cite{IC&SRN_SW01}.
}
\label{fig:similarity_200/200}
\end{figure}

\begin{figure}[htbp]
\centerline{
\resizebox{0.9\textwidth}{!}
{\includegraphics[clip=]{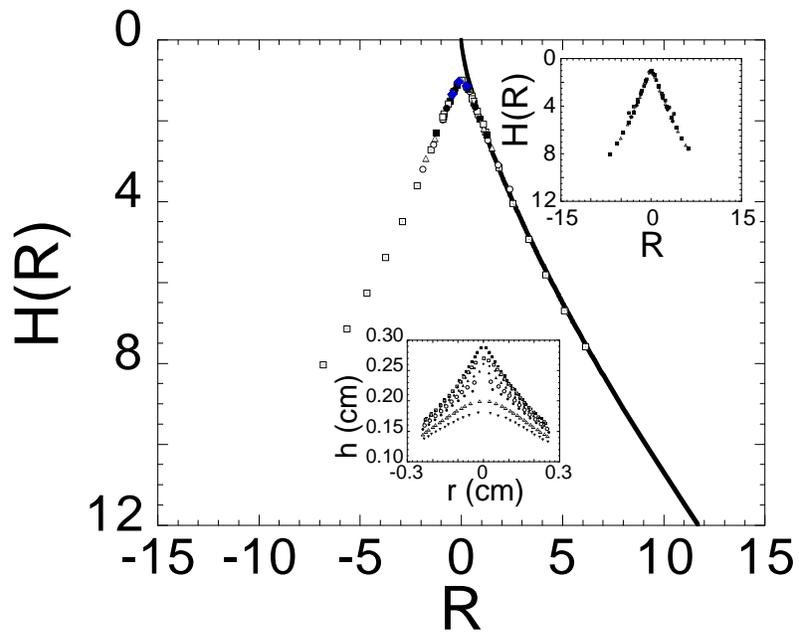}}}
\caption[The scaled hump profiles for system 5]{
The scaled hump profiles for system 5.  The lower inset shows eight profiles taken 
from the $S = 0.667$ data set.  The main figure shows the same profiles after scaling.  The 
solid line corresponds to a power law of the form $R^{0.72}$.  In the upper inset we compare 
the universal curves for the $S = 0.984$ cm, $0.921$ cm, $0.889$ cm, $0.667$ cm, $0.445$ cm data 
sets.
}
\label{fig:similarity_200/1}
\end{figure}

Here $h(r)$ is the hump profile and the value of $h_{c}$ is taken from the scaling relations.  
This transformation shifts and scales the profiles so that the singularity is located at the 
origin and the maximum hump heights are located at $H = 1$ and $R = 0$.  Figure \ref{fig:similarity_200/200} 
reproduces the similarity analysis of Cohen and Nagel (system 6) for the hump profiles 
near the transition while figure \ref{fig:similarity_200/1} shows the results of an identical treatment performed 
on data taken for the small $\nu_l$ system (system 5).  

Figures \ref{fig:similarity_200/200}  and \ref{fig:similarity_200/1} show a series of eight 
scaled profiles for the $S = 0.830$ cm 
data set in the large $\nu_l$ system and the $S = 0.667$ cm data set in the small $\nu_l$ system.  The 
bottom insets in the figures show an overlay of the different hump profiles which are 
scaled in the main figures.  Both figures display excellent collapse for the hump profiles.  
The solid lines are power-laws which fit the scaled profiles in the region beyond the 
parabolic hump tips.  The scaling relation in figures  \ref{fig:Scaling200/200} 
and  \ref{fig:Scaling200/1} can be used to predict 
the exponent, $x$, in the power-law fits.  Inserting the observed scaling dependence  
$\frac{h_c-h_{max}}{h_{max}} = (\frac{\kappa}{n})^\beta$ into the observed form $H(R) = R^x$ which fits the profile, the 
following relation is obtained:
\begin{equation}
\label{exponent_predict}
\frac{h_c-h(r)}{h_{max}} (\frac{\kappa}{n})^{-\beta} = r^x (\frac{\kappa}{n})^x.
\end{equation}
Since, for a given $r$, the functions $\frac{h_c-h(r)}{h_{max}}$ and $r^x$ have 
constant values, $x$ must 
equal $-\beta$.  More intuitively, as $Q$ is increased, the parabolic tip regions decrease their 
radial length scale and are simultaneously pulled towards the singularity in the axial 
direction leaving behind power-law profiles with exponents that reflect the rate at which 
these length scales are decreasing.  The fits to the profiles indicate that $x = 0.72 \pm 0.08$ 
and $x = 0.72 \pm 0.10$ for the high and low $\nu_l$ systems respectively. Both of these 
exponents are within error (although slightly 
smaller) of the exponent observed in the scaling relation of figures  \ref{fig:Scaling200/200} 
and  \ref{fig:Scaling200/1} which, 
respectively, predict values of $0.85 \pm 0.09$ and $0.86 \pm 0.10$ for $x$.  

Typically, the observed scaling dependencies in these types of problems result from
the local stress balance.  A scaling analysis where the viscous stresses of the upper and 
lower fluids balance the stress arising from the interfacial curvature predicts 
linear scaling dependencies and conical profile shapes.  The non-linearity of the observed 
dependencies and the lack of dependence of the similarity solution on $\nu_{l}$ indicate 
that a different stress balance may govern the flows (e.g. only viscous stress due to 
upper fluid balances stress due to the interface curvature). Furthermore, non-local effects 
coupling themselves into the solution could account for 
the slight differences between the values of the exponents $\beta$ and $x$. Further 
evidence for the existance of these coupling effects is presented in section \ref{sec:conclusion}.

In both figures, the upper right inset shows a comparison of the similarity 
solutions for five different tube heights.  The profiles corresponding to the different tube 
heights all display the same power-law dependence.  Within error, $\frac{k}{n}$ (taken from figures 
 \ref{fig:Scaling200/200} and  \ref{fig:Scaling200/1}) can be used to scale the radial 
components of these profiles and obtain good 
collapse.  Recall that the normalization prefactors $n(S)$ decrease roughly as $\exp^{-2.5 S}$.  
Here, this decrease is correlated with the observation that the profiles become shallower 
at larger $S$.  The points of deviations for the $S =0.255$ cm and $0.381$ cm profiles in the 
large $\nu_l$ system, and the $S=0.445$ cm profile in the small $\nu_l$ system, mark the 
transition from the similarity regime to the matching regime beyond which the profiles 
become asymptotically flat.  At large enough radii all of the scaled profiles display these 
deviations. 

\begin{figure}[htbp]
\centerline{
\resizebox{0.7\textwidth}{!}
{\includegraphics[clip=]{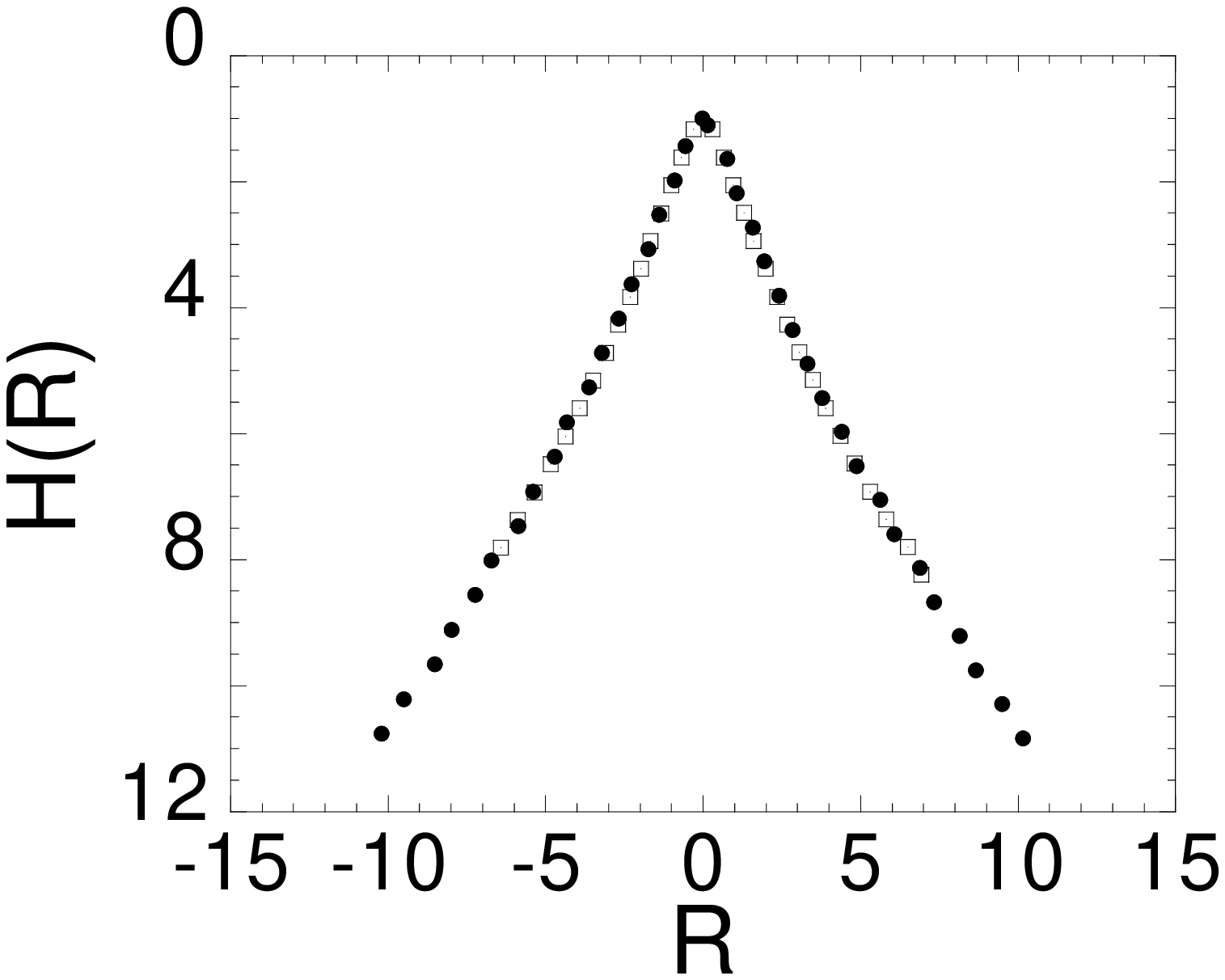}}}
\vspace{5mm}
\centerline{
\resizebox{0.73\textwidth}{!}
{\includegraphics[clip=]{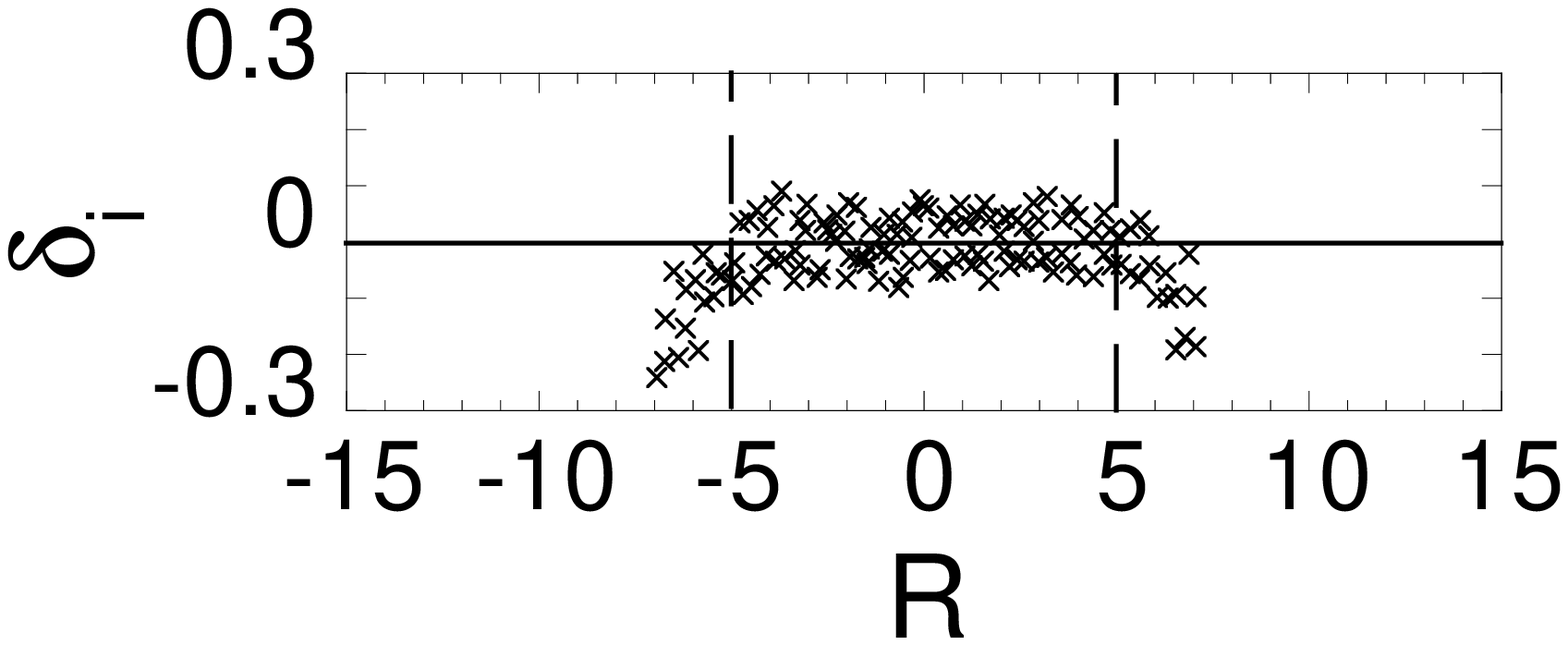}}
}

\caption[Comparison of similarity profiles for systems 5 and 6]{
Comparison of the similarity profiles for systems 5 and 6.  The closed symbols correspond 
to the $S=0.830$ cm data for system 6 while the open symbols correspond to the 
$S=0.667$ cm data for system 5. Figure \ref{fig:similarity_200/200_200/1}b shows 
a plot of residual quantity, $\delta_i$ for all of the data points. Here, $\delta_i$ 
is the minimum distance between each point on the system 6 curve and the value 
of a local linear fit to the system 5 curve at the same R value. The region between the vertical dotted 
lines corresponds to the similarity regime where the residuals become centered 
around zero. The $\chi^2$ calculation was performed within this similarity regime.
}
\label{fig:similarity_200/200_200/1}
\end{figure}

As a final check, figure  \ref{fig:similarity_200/200_200/1} shows a comparison of the 
similarity profile for 
$S=0.831$ cm in the large $\nu_l$ system and the similarity profile for $S=0.667$ cm in the 
small $\nu_l$ system.  An error analysis calculation\footnote{Since the 
data points for both curves were not aligned, $\chi^2$ was calculated in the following way: 
First a local linear fit was used to interpolate the value of curve$_1$ between the data points.  Second, 
the minimum distance between the interpolated curve and the value of the points in 
curve$_2$, $\delta_{i,2}$ was calculated. 
The $\chi^2$ value for the entire curve is defined as:
$\chi^2 = \frac{1}{N} \sum \frac{\delta_{i,2}^2}{\sigma_{1x}^2 + \sigma_{1y}^2 +(\sigma_{2x}^2 + \sigma_{2y}^2)/n}$
where $N$ is the number of points compared, $\sigma_{1x}$ and $\sigma_{1y}$ are the experimental 
errors associated with the $x$ and $y$ points in curve$_1$, $\sigma_{2x}$ and $\sigma_{2y}$ 
are the experimental errors associated with the $x$ and $y$ points in curve$_2$, n is equal to the number 
of points used in making the linear fit to curve$_1$, and the sum is taken over the 
index $i$ which labels the points in curve$_2$.} for the data points located between 
$-5 < R < 5$ (where the residuals are centered around zero - see figure 
\ref{fig:similarity_200/200_200/1}) shows that $\chi^2 \approx 1.6$. This $\chi^2$ value indicates 
that the differences between the two similarity curves are on the same order 
as the experimental uncertainty which is taken to be a quarter of a scaled pixel.  
The excellent collapse verifies that $\nu_l$ and the viscosity ratio do not affect the similarity 
solution and emphasizes that the scaling behavior observed in these systems is robust. 

\section{Mapping Out The Transition Location}\label{sec:SuvsQ}

Having shown that the viscosity ratio does not affect the scaling relations and 
self-similarity characterizing the detailed structure of the transition, I return 
to the question of whether $S_{u}$ is affected by such a change in parameters.
There is a large body of work dating back to the late 1940's 
(see, for example, \cite{Rouse56,Gariel49,Craya49,Muskat49}) which addresses 
the mapping out of the selective withdrawal transition location.  
A large portion of this work focuses on the 
problem of extracting crude oil deposits without withdrawing any of the water which is 
often trapped beneath the oil.  Nearly all of the experimental studies found in the 
literature which address these large scale extraction problems assume that the 
flows have a high Reynolds number or, equivalently, that the viscous stresses are 
negligible.  Therefore, low viscosity fluids were used to model the flows.  One exception 
arises in the modeling by \cite{Blake&Ivey86} of magma layer mixing during volcanic eruptions.  
However, as described earlier, miscible fluids were 
used in those investigations.  Since, as \cite{JRL89} showed, the absence of 
surface tension allows for the 
withdrawal of both fluids at any withdrawal rate, these particular studies never 
addressed the actual selective withdrawal transition.  With the advent of new technologies 
which use the selective withdrawal geometry in conjunction with low Reynolds number 
flows (see, for example, \cite{Itai_coating,Ganan-Calvo98}), it has become increasingly 
important to determine the effect of the viscous stresses on the transition. 

\begin{figure}[htbp]
\centerline{
\resizebox{0.9\textwidth}{!}
{\includegraphics[clip=]{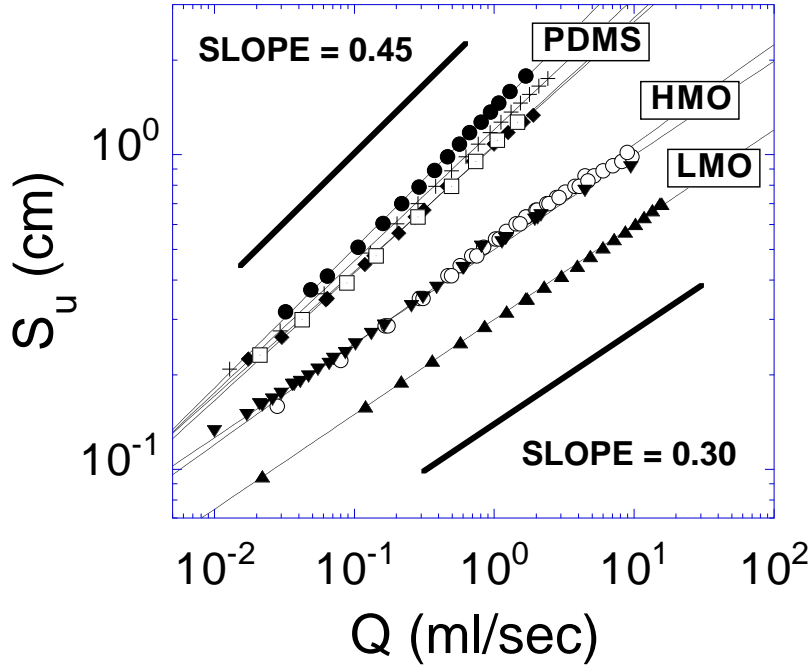}}}
\caption[Plots of $S_u$ versus $Q$ for systems listed in table \ref{table:System_properties}]
{Plots of the transition tube height $S_u$ as a function of $Q$ for the 
seven systems listed in table \ref{table:System_properties}.  Table \ref{table:System_properties} lists 
the symbol used in plotting the $S_u$ curve for each system.  For each system, it is observed that $S_u \propto Q^{\alpha}$ 
where the power $\alpha$ (listed in table \ref{table:System_properties}) ranges between 0.30 and 0.45.
The $S_u$ curves cluster into three groups which correspond to the three different upper 
fluids used in the experiments. Each cluster is labeled with the corresponding upper fluid. 
}
\label{fig:Su_vs_Q_compare}
\end{figure}

Figure \ref{fig:Su_vs_Q_compare} shows a plot of $S_u$ versus $Q$ 
for seven pairs of fluids.  Table \ref{table:System_properties} lists the fluids used in making 
these measurements along with the measured values of the interfacial tension, $\gamma$, the fluid 
densities, $\rho_u$ and $\rho_l$, the density mismatch, $\Delta\rho$, the fluid viscosities, 
$\nu_u$ and $\nu_l$, and the viscosity ratio $\nu_l/\nu_u$.  

There are a few obvious trends.  First, all of the data sets show that $S_u \propto 
Q^{\alpha}$ where the power $\alpha$ ranges between 0.30 and 0.45. The last row in Table 
\ref{table:System_properties} shows the value of the exponent for the power-law used to 
fit the $S_u$ curves for each system. Second, for a given value of $Q$, an increase in $\nu_u$,
the upper fluid viscosity, increases $S_u$.  The three distinct clusters of  curves in 
figure \ref{fig:Su_vs_Q_compare} correspond to three distinct 
upper fluid viscosities.  As figure \ref{fig:Su_vs_Q_compare} shows, a five fold increase 
in $\nu_u$ can increase $S_u$ by a factor of two.  Third, for a given value of $Q$, even a 
thousand fold increase in the lower fluid viscosity does not significantly affect $S_u$.  These 
last two observations indicate that it is the upper fluid viscosity rather than the viscosity 
ratio which affects the transition location within the $S$ vs. $Q$ parameter space.  

The remaining trends in the data are weak.  In order to ascertain the 
effects of the surface tension, $\gamma$, and density mismatch, $\Delta \rho$ on the 
transition straw height, these 
parameters must be varied by large amounts.  Unfortunately, reducing $\gamma$ 
below about 20 dynes/cm causes mixing of the fluid layers and results in a diffuse 
interface under high shear rates.  Therefore it is very difficult to vary $\gamma$ 
in the experiments by more than a factor of two.  Another difficulty arises from an 
inability to decouple the changes in these parameters for different systems.  
For example, when comparing
systems 1 and 4, an order of magnitude increase in $\Delta \rho$ is coupled with a factor of 
two decrease in $\gamma$.  Since both $\Delta\rho$ and $\gamma$ act to stabilize the interface, 
the effects due to these variations may 
cancel.  A further complication arises due to the effect of surfactant 
concentration on the interface.  As will be discussed in Appendix A, the 
accumulation of surfactants on the interface tend to shift $S_u$ by 
an amount which is comparable to the slight shifts seen in top four curves (which form 
the uppermost cluster) in figure \ref{fig:Su_vs_Q_compare}.  More careful measurements 
of $S_u$ are necessary in order to resolve these weaker trends.

It is instructive to compare these results with currently available theoretical 
predictions for the transition which take into account both the viscous stresses 
and the surface tension.  While there are no current numerical studies which have been 
tailored to address this exact problem, the closest approximation to such a treatment 
can be found in the \cite{JRL89} paper on selective withdrawal for zero Reynolds number 
flow.  While his simulations were performed for equal viscosity fluids, the
fact that the lower fluid viscosity is irrelevant to the transition location can be 
used to compare the simulations with the experiments for which the fluid 
viscosities are unequal. Since those simulations were designed to address 
large scale magma flows rather 
than the small scale fluid flows found in the experiments, care must be taken in making a 
comparison of the results.  For example, the simulations were designed to model a system 
with a point sink which is located many capillary lengths (defined as $\sqrt{\frac{\gamma}{\Delta \rho g}}$) away 
from the interface.  However in the experiments, $S$ is comparable to the capillary 
length.  Thus, the experiments are performed at much smaller values of $Q$ and $S_u$ 
than the simulations making it impossible to quantitatively compare the results.  
Nevertheless, since it is expected that $S_u$ is a smooth function of $Q$, one 
can check that the two data sets are consistent with this expectation. 

Another issue which needs to be addressed is the determination of the Reynolds 
number for the experimental flows. There are a variety of ways in which the Reynolds number 
can be defined.  The greatest change in the velocity of the upper fluid occurs along the line 
connecting the tip of the hump and the withdrawal tube.  Since the fluid at the tip of the 
hump is stationary the magnitude of this change in velocity is $\frac{Q}{\pi(D/2)^2}$.  The 
distance of the tube orifice to the interface is the largest length scale affecting the 
flows\footnote{For this range of flow rates and density mismatch values, the capillary 
length scale is a little smaller than $S$.  However, it is conceivable that for a set of 
density matched fluids or for measurements at very small $S$ the capillary length could 
become larger than $S$.}.  Therefore, the quantity $Re_{max} = \frac{Q S}{\pi(D/2)^2 \nu}$ 
can be used as an upper bound for the Reynolds number.  In figure \ref{fig:Su_vs_Q_compare}, 
the upper most cluster of $S_u$ curves has $0.005 < 
Re_{max} < 6$, the middle cluster of curves has $0.03 < Re_{max} < 50$, and the lowest 
curve has $0.05 < Re_{max} < 350$.  These estimates imply that a comparison with results 
from a zero Reynolds number simulation must be 
treated with caution.  For example, the effects of the fluid's inertia could manifest 
themselves as a change in $\alpha$, the value of the exponent for the power-law fits 
in figure \ref{fig:Su_vs_Q_compare} \footnote{W. W. Zhang private communication.}.  Such 
an explanation could account for the decrease in $\alpha$ which occurs when 
$\nu_u$ is decreased from 10 St to 1.9 St.  However, since an equally large change in 
Reynolds number can occur along the $S_{u}$ versus $Q$ curves it is curious that kinks 
which may be indicative of such a change are not observed.  
 
\begin{figure}[htbp]
\centerline{
\resizebox{0.9\textwidth}{!}
{\includegraphics[clip=]{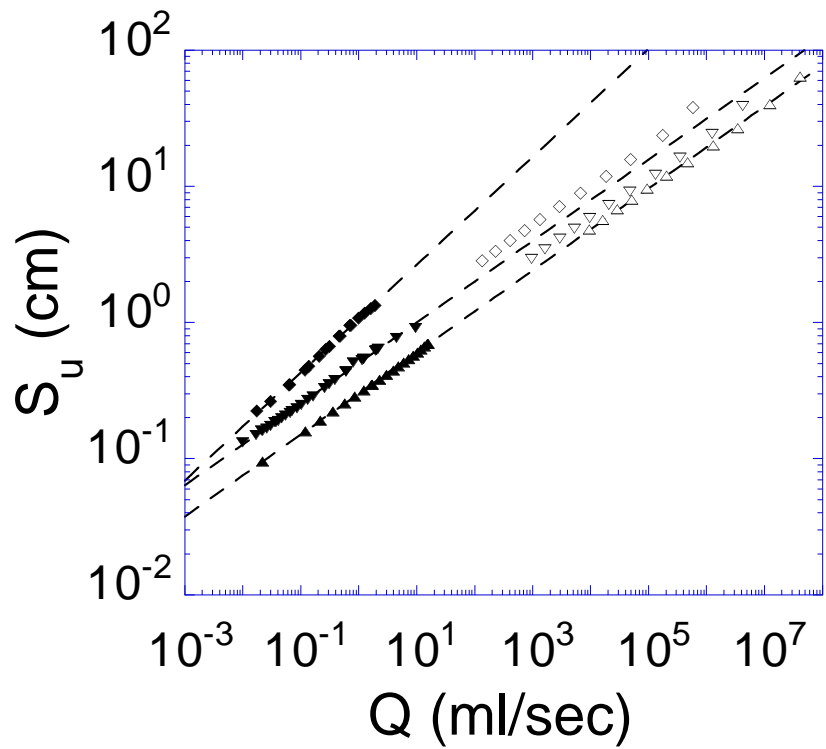}}}
\caption[Comparison of experimentally measured $S_u$ versus $Q$ curves with those 
predicted by simulations of Lister]
{ 
Comparison of the experimentally measured $S_u$ curves for systems 4, 6, and 7 in 
Table \ref{table:System_properties} with those predicted by the simulations of \cite{JRL89}.  
The closed symbols depict the experimental measurements while the corresponding open symbols depict the simulation 
predictions for a system with the same fluid parameters.  The dashed lines are power-law fits to the experimental data. 
}
\label{fig:Su_vs_Q_theory_compare}
\end{figure}

In figure \ref{fig:Su_vs_Q_theory_compare} the $S_u$ vs. $Q$ curves are compared 
with the predictions of Lister for systems with $\nu_u =  10$ St, $\nu_u = 1.9$ St, 
and $\nu_u = 0.57$ St corresponding to systems 4, 6 and 7 in table \ref{table:System_properties}.  
The solid and open symbols denote data from the experiments and simulations respectively.  
The dashed lines represent power-law fits to the experimental results.  
These fits are projected into the regime for which the simulations have a prediction for the 
value of $S_u$.  Since Lister's simulations were performed using dimensionless variables, the 
proper values for the fluid parameters were used to determine the values of $S$ and $Q$.  
Furthermore, the simulation flow rate is reduced by a factor of two in order 
to make a first order correction for the fact that the simulations use a point sink rather 
than a tube to withdraw the fluids.

There are two features of the comparison which need be addressed.  First, the 
simulations curves for the $\nu_u = 10$ St system fall below the projected power-laws.  
In Section \ref{sec:trans_struct} evidence 
was provided for the lack of dependence of $S_u$ on the straw diameter $D$ for the range of flow 
rates used in the experiments.  Therefore it is unclear whether geometric factors alone 
could account for the large shift which is necessary for the collapse of the numerical 
results onto the projected experimental power-laws.  Second, it is disturbing that the 
power-law exponents for the systems which use less viscous (rather than more viscous) 
upper fluids agree more favorably with the zero Reynolds number simulation results.  As stated above, 
the shift in the power-law exponent for the experimental results indicates that there may 
be more than one way to balance the stresses acting on the interface.  The simulation 
results only address the regime 
where surface tension and buoyancy effects play an equal role in stabilizing the interface 
and therefore do not observe the change in the $S$ vs. $Q$ power-law exponents\footnote{W. W. Zhang private communication.}.  Nevertheless, the 
fact that the simulations predict a power law exponent of one third and the fact that 
the experimentally measured exponents are centered around this value deserves notice.  

Finally, it is often possible to use the scaling relations observed in systems 
approaching a singularity to help identify which stresses are balancing each 
other.  Such an identification for the different fluid 
systems could in turn help explain the observed changes in the exponent $\alpha$.

\section{Conclusions}\label{sec:conclusion}

	In this paper, it has been shown that the transition 
straw height for a given flow rate, $S_u$, changes as: $S_u \propto Q^\alpha$, where $\alpha$ ranges 
between about 0.45 and 0.30.  This is the first instance in which experiments comparing $S_u$ for 
different systems of immiscible fluids with flows in the low 
Reynolds number regime have been performed.  The data indicates that for the selective 
withdrawal system, the upper fluid viscosity rather than the viscosity ratio determines the value of $S_u$. 
It has been shown, that when the withdrawal tube is sufficiently close to the 
interface, the transition is hysteretic: the straw height at which the spout collapses 
back into a hump is different from $S_u$.  At larger 
straw heights, the difference between $S_u$ and the spout to hump transition straw height, 
$\Delta $S, decays exponentially.  This effect has been linked to the tube diameter which at low 
$S$, sets a length scale for the final mean radius of curvature $1/\kappa_u$.  However, the transition 
remains discontinuous and $\kappa_u$ remains finite even at large $S$.

I have performed a detailed comparison of the 
scaling relations for two systems which have the same upper fluid viscosity but have 
a lower fluid viscosity which is different by a factor of two hundred.  The surface tension 
for the two systems was different by about 13$\%$ while the density mismatch was 
different by a factor of three (See systems 5 and 6 in table \ref{table:System_properties}).  It is observed that 
up until the cutoff, the hump profiles behave as though they are approaching a singular 
solution where, at the flow rate $Q_c$ the hump height would be equal to $h_c$ and the mean curvature, $\kappa$, 
would diverge.  The quantity  $\frac{h_c-h_{max}}{hmax}$ has been shown to scale as $(\frac{k}{n})^\beta$ where $\beta 
= -0.85 \pm 0.09$ for the high $\nu_l$ system and $\beta = -0.86 \pm 0.10$ for the low $\nu_l$ system.  
These scaling relation were used to collapse the hump profiles for different flow rates and 
straw heights near the transition onto a single universal curve.  The region of the similarity 
profiles located beyond the parabolic tip can be fit with a power law which has an 
exponent $x = 0.72$ for both systems. The results show that both the 
scaling exponents and the shape of the similarity solution 
are independent of $\nu_l$ and the viscosity ratio.
In fact, a direct comparison of the similarity solution for both fluid systems indicates 
that, within error, the curves are identical. 

As described in the introduction, in the 2-D analogue to the selective withdrawal 
problem the lower fluid viscosity plays a fundamental role in determining how close the 
system can approach the cusp singularity. In the 3-D selective withdrawal system,  
the asymptotic value of $\kappa_u$ shows little or no dependence on $\nu_l$.  
Therefore, the effects of the lower fluid viscosity must 
enter as higher order terms which are undetected by the experiments.  
Since $\kappa_u$ is independent of the flow rate at large $S$,  an increase in the 
Capillary number, $\nu_u \frac{Q}{r^2}/\gamma$ can also be ruled 
out as a method of getting the system closer to the singularity.  

The saturation value of the mean curvature, $\kappa_{usat}$ does show some 
dependence on the upper fluid 
viscosity $\nu_u$.  For example, systems 5 and 7 in table \ref{table:System_properties} 
show that when $\gamma$ and $\Delta \rho$ 
are kept constant, decreasing $\nu_u$ from 2.0 St to 0.57 St, decreases $\kappa_{usat}$ by a factor of six.  
However, as systems 2 and 5 in the table indicate, this trend is not uniform.  
Shallow hump profiles are also observed for nearly inviscid systems which use air as the 
upper fluid and water as the lower fluid.  Therefore, it is likely that one of the ways in 
which $nu_u$ can affect $\kappa_{usat}$ is by shifting the balance between the viscous and inertial 
terms in the governing Navier-Stokes equations.  For the remaining systems which have a 
lower Reynolds number characterizing the flows, even a factor of five increase in $nu_u$ does 
not significantly affect $\kappa_{usat}$.  

Since at low enough Reynolds numbers the 
fluid viscosities do not affect $\kappa_{usat}$, the only remaining parameters which may act to set 
the length scale for the cutoff are the density mismatch $\Delta \rho$ and surface tension $\gamma$.  With 
the exception of systems 1 and 6, the large error in the data makes it difficult to 
distinguish between the $\kappa_{usat}$ values for the different systems.  Also, the trends in the 
data are too weak to allow for a determination of the effects of $\gamma$ on $\kappa_{usat}$.  
However, a comparison of system 1 with systems 2, 3, and 4 and a comparison 
of system 6 with system 5 indicates that an increase in $\Delta \rho$ results in a higher 
value for $\kappa_{usat}$. Note that for both comparisons $\nu_u$ is kept constant.  
This result is intriguing since the similarity treatment for the profiles is localized to the hump tip 
whereas the effects of the density mismatch should only affect the system on a length 
scale comparable to the capillary length $\sqrt{\frac{\gamma}{\Delta \rho g}}$.  These observations suggest that 
the matching region which connects the profile near the tip of the hump to the flat 
interface at large radii is responsible for setting the length scale for the cutoff \footnote{J. Eggers, W. W. Zhang, S. R. Nagel, J. Wyman, and H. A. Stone (private communication).}.  
Such an effect could also explain the slight difference 
between the observed value of the power-law exponent describing the similarity solution($x$) 
and the value predicted by the scaling relations($\beta$).  The importance of identifying which 
parameters determine the length scale for the cutoff warrants a more careful investigation 
of the $k_{usat}$ dependence on $\Delta \rho$, $\gamma$ and the local interfacial boundary conditions. 
	
The robustness of the similarity analysis shows that singularities can be 
used to organize the study and classification of the steady state hump profiles near the 
selective withdrawal transition.  In particular, the discontinuous nature of the transition, marked by the cut-off curvature $\kappa_{u}$, 
coupled with the display of scaling behavior suggests a transition structure which is 
remarkably similar to that of weakly-first-order thermodynamic transitions.  Whether this 
analogy hints at some deeper relationship between the classification schemes for weakly first order
thermodynamic transitions and those for these types of topological transitions remains to 
be shown.

I am grateful to S. R. Nagel, W. W. Zhang, S. Venkataramani, J. Eggers, H. A. Stone, 
D. Mueth, J. Wyman, T. J. Singler, H. A. Lyden, J. N. Israelachvili, C. C. Park, 
S. Chaieb, V. Putkaradze, R. Parthasarathy, S. N. Coppersmith, T. A. Witten, 
L. P. Kadanoff, P. Constantin, R. Scott, E. Blucher, T. Dupont, H. Diamant, and V. C. Prabhakar 
for sharing their insights and their help with editing this manuscript.  
This research is supported by the University of Chicago 
(MRSEC) NSF DMR-9808595 and NSF DMR-0089081 grants.
%
%
\appendix
\section{Surfactant effects}\label{sec:appendix}

Surfactants are chimeric molecules consisting of hydrophobic and hydrophilic 
sections which tend to aggregate at fluid interfaces\footnote{For some oil-oil 
(for example Polybutadiene and PDMS) interfaces the random thermal 
energy is able to keep the surfactants off the interface.  However 
these systems usually have a very low surface tension and therefore 
develop diffuse interfaces under high shear rates.}. Even under 
very controlled conditions, it is difficult to keep fluid systems free of surfactants 
for long periods of time.  Since, surfactants can significantly affect the interfacial 
tension and surface flows, part of understanding the details of any fluid interface 
problem entails isolating and accounting for the effects of surfactants.

In the present studies there are two types of effects which are thought to result 
from the presence of surfactants.  First, for a given flow rate, it is observed that over a 
period of several days, $S_u$ increases by a small amount indicating that the surface 
tension in the system may have been uniformly reduced.  Second, at the transition, the 
system oscillates between the hump state and the spout state over a period of about 
minute\footnote{Note that the time scale for this effect is much larger than the time scale for the pump 
induced noise in the withdrawal rate which can produce a similar effect when 
the system is very close to the transition.} in a manner similar to that observed by \cite{Bruijn93} 
in tip streaming of drops under 
shear flow.  In the tip streaming experiments, a straining flow stretches a drop of fluid
so that the two tips of the drop (each of which is analogous to the hump tip) 
have a high curvature. At sufficiently high shear 
rates the drop can enter the streaming state (analogous to the spout state) where a small jet of fluid  eminates 
from each of the drop tips.  It is observed that near the transition the drops oscillate between
the stretched drop state and the streaming state. This oscillatory behavior has been linked to the 
accumulation of surfactants near the stagnation point located at the drop tip.  
This accumulation lowers the local surface tension and causes a transition 
to the streaming state.  The jet ``sweeps'' the interface, reduces the local concentration of 
surfactants, and subsequently collapses back into the drop state at which point the cycle 
begins again.  While the local boundary conditions for this problem are different from 
those of the selective withdrawal problem, the 
flow patterns are remarkably similar indicating that the oscillatory behavior observed in 
both systems may be correlated.  Below, I explain how both of these effects influence the 
results presented in the main body of this paper.

\begin{figure}[htbp]
\centerline{
\resizebox{0.7\textwidth}{!}
{\includegraphics[clip=]{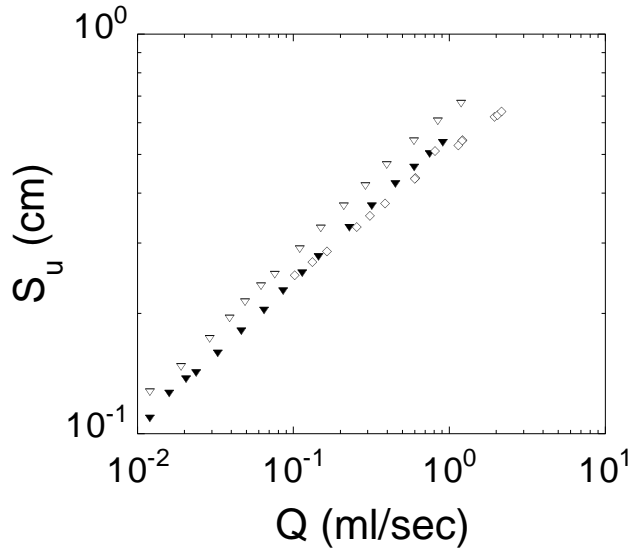}}}
\caption[Plots of $S_u$ versus $Q$ for system 6 
in Table \ref{table:System_properties}]{
Plots of the transition tube height $S_u$ as a function of $Q$ for system 6 
in Table \ref{table:System_properties}.  The filled down-triangles correspond to 
measurements taken just after the interface was cleaned. 
The open down-triangles correspond to measurements taken after the surfactant 
concentration at the interface was allowed to equilibrate over a period of a week. 
The open diamonds correspond to measurements taken after the system was cleaned once again.  
}
\label{fig:Su_vs_Q_clean/dirty}
\end{figure}

\begin{figure}[htbp]

\centerline{
\resizebox{0.7\textwidth}{!}
{\includegraphics[clip=]{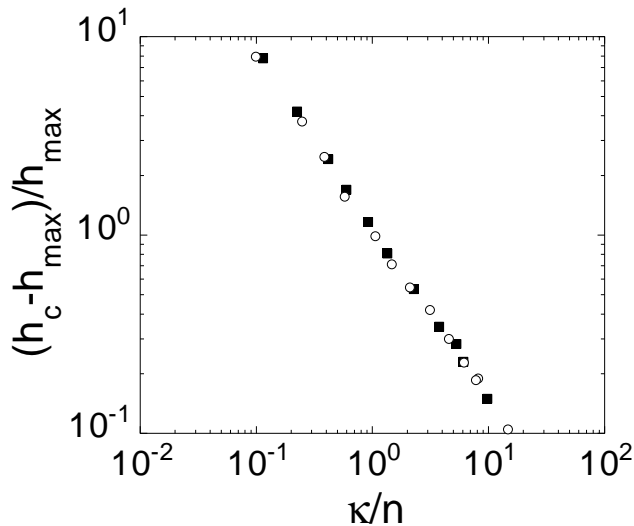}}}

\caption[Plots of $\frac{h_c-h_{max}}{h_{max}}$ versus $\frac{\kappa}{n}$ for system 
6 in Table \ref{table:System_properties}]
{Plots of $\frac{h_c-h_{max}}{h_{max}}$ vs. $\frac{\kappa}{n}$ for system 
6 in Table \ref{table:System_properties}. The squares correspond to 
measrements taken after the surfactant concentration at the interface 
was allowed to equilibrate over a period of a week. The open circles 
correspond to measurements taken after the system had been cleaned.
}
\label{fig:scaling_clean/dirty}
\end{figure}

Figure \ref{fig:Su_vs_Q_clean/dirty} shows plots of $S_u$ vs. Q for three data sets corresponding to 
measurements performed on system 6.  The filled down-triangles correspond to 
measurements taken just after the fluid interface was cleaned\footnote{When the 
system is in the spout state, surfactants, which have accumulated on the interface 
are ``swept'' into the straw and deposited in the waste container. By leaving the 
system in the spout state for long periods of time, the surfactant concentration 
over the entire interface can be significantly reduced.}. 
The open down-triangles, correspond to measurements of $S_u$ taken after the surfactant 
concentration at the interface was allowed to equilibrate for a period of a week. A uniform 
20$\%$ increase in $S_u$, or, equivalently, a 50$\%$ decrease in the transition flow rate 
is observed. Upon cleaning the interface once again, the $S_u$ data 
points return to their original value (open diamonds).  Note that the observed shift in 
$S_u$ resulting from the differing surfactant concentration is quite small compared with 
the shift observed when $\nu_u$ is increased by a factor of five.  Nevertheless, the fact that the transition 
flow rate can change by nearly 50$\%$ depending on the surfactant concentration is 
noteworthy.  Finally, in figure \ref{fig:scaling_clean/dirty}, a comparison between the $\frac{h_c-
h_{max}}{h_{max}}$ vs. $\frac{\kappa}{n}$ curves for the equilibrated system (squares) and the 
clean system (open circles) indicates that the shift in $S_u$ does not affect the scaling relations.

While the increase in the $S_u$ values is most likely due to a uniform reduction in 
$\gamma$ over the entire interface, the dynamic nature of the observed hump to 
spout oscillations implies that (as with tip streaming) the flows may cause local variations in the 
surfactant concentration.  In this picture, when the system is in the hump state, 
surfactants are dragged towards the hump tip by the surface flows.  
If the system is sufficiently close to the transition, this accumulation, which 
lowers the surface tension locally, causes the hump to increase its height 
and curvature and ultimately drives the system into the spout state. 
Once in the spout state, the local surfactant concentration reduces over time 
and the system eventually decays back into the hump state.  The detailed characteristics 
of this cyclical behavior depend on the surfactant concentration over the entire interface. 

\begin{figure}[htbp]

\centerline{
\resizebox{0.7\textwidth}{!}
{\includegraphics[clip=]{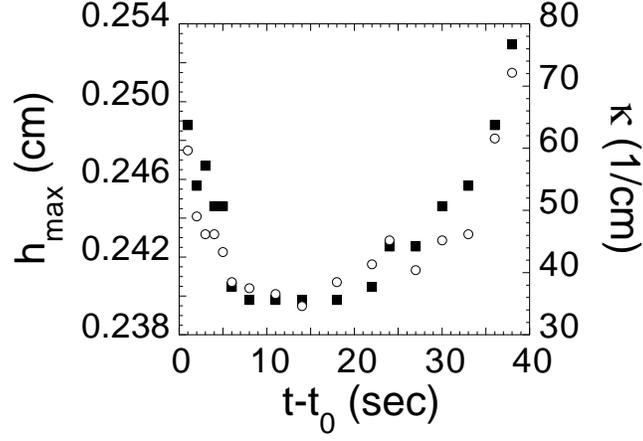}}}
\vspace{10mm}
\centerline{
\resizebox{0.7\textwidth}{!}
{\includegraphics[clip=]{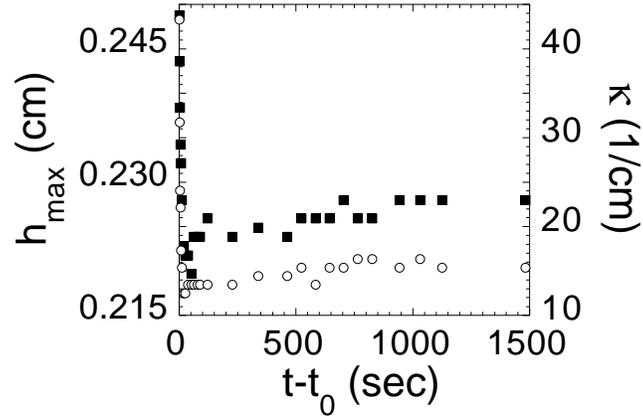}}}
\caption[Plots of the time dependence for $h_{max}$ (squares) and $\kappa$ (open circles) just 
after spout collapse]
{Plots of the time dependence for $h_{max}$ (squares) and $\kappa$ (open circles) just 
after spout collapse for system 3 in Table \ref{table:System_properties}. The figures 
correspond to measurements taken at constant $S$ but at different $Q$s. In both 
figures, $t_0$ corresponds to the time at which the spout collapses into a hump. 
Figure \ref{fig:h_k_time_dep}a shows measurements for a system which is very close 
to the transition with $\frac{Q_c - Q}{Q} \approx 0.004$. Figure \ref{fig:h_k_time_dep}b 
shows measurements for a system which slightly farther away from the transition 
with $\frac{Q_c - Q}{Q} \approx 0.01$. 
}
\label{fig:h_k_time_dep}
\end{figure}

When the entire interface has been cleaned, it takes a long time for the surfactants 
to accumulate in concentrations large enough to affect the steady state.  Therefore, the increase 
in the hump curvature and height is slow. Furthermore, once the interface forms a spout, 
it takes a very short amount of time for the local concentration of surfactants to be reduced.
Consequently, systems that have been cleaned spend a very small fraction of the hump to spout 
oscillation cycle in the spout state. Since many of the measurements discussed in the paper 
were performed in the vicinity of the transition, care needed to be taken when measuring 
the hump height and curvature. In order to understand how $h_{max}$ and $\kappa$ 
change with time, the following experiment was performed.  
First, the system was placed in the hump state.  Then, the flow rate was 
increased momentarily so that the interface formed a spout for a short period of time.  Just after spout 
collapse, the time dependence of the hump curvature and height was measured. Figures 
\ref{fig:h_k_time_dep}a and \ref{fig:h_k_time_dep}b show the results of measurements performed for the same system, 
with the same value of $S$, but at different values of $Q$.  The initial decay, which occurs over a 
time scale of about five seconds, gives some measure of the relaxation time for the flows 
in these systems.  Following this initial decay, the height and curvature values plateau. 
These plateaus are associated with a regime where the local surfactant concentration is 
too low to affect the shape of the interface.  
When the system is near the transition so that $\frac{Q_c-Q}{Q} \approx 0.004$, the 
plateau regime is short lived (about fifteen seconds) and is followed by a regime in which 
both the height and curvature increase their values (figure \ref{fig:h_k_time_dep}a). 
However, when the experiments are performed at larger values of  $\frac{Q_c-Q}{Q}$ the 
plateau regime is longer.   Figure \ref{fig:h_k_time_dep}b shows that when 
$\frac{Q_c-Q}{Q} \approx 0.01$ the plateau regime lasts for over twenty minutes.

When collecting data for the scaling relations, it is important to determine the 
value of the hump height and mean curvature in the plateau regime.  Also, a larger amount 
of error must be assigned to data points taken from plateau regimes that are short lived.  
Note however that since the plateau regime lengthens quite rapidly as $\frac{Q_c-Q}{Q}$ 
is increased, for the cleaned interface, these precautions only apply to the one or 
two data points in the scaling relations, which are closest to the transition.

While all of the scaling data presented in the main body of this paper was taken for 
clean interfaces, it is useful to understand the changes in the transition structure 
which occur when the surfactant concentration has been allowed to equilibrate over 
a period of days. For equilibrated interfaces the increase in the hump height and 
curvature occurs over a shorter amount of time so that the plateau regimes shorten. 
Therefore, greater caution needs to be taken in the measurements of the scaling relations.
However as figure \ref{fig:scaling_clean/dirty} shows, when such precautions are 
taken no noticeable changes in the scaling relations are observed.

Another characteristic of the equilibrated systems, is that (unlike the cleaned systems) 
they can spend a large fraction of the hump to spout oscillation cycle 
in the spout state. During the time which the system spends in the spout state, 
it is possible (even at high Q or S) to quickly raise the straw and cause the spout to collapse back into 
the hump state\footnote{Since the spout drains the lower fluid, the straw height $S$ increases with time. 
Therefore, it is necessary to check that the amount of hysteresis is actually larger than the increase 
in $S$ which occurs during the time it takes to perform the measurement.}. The amount $\Delta S$ 
necessary to cause the collapse can be measured. I find that the value of $\Delta S$ varies depending on 
conditions.  For example, if the $\Delta S$ measurement is performed just 
after the local concentration of surfactants has been reduced, only a small amount of 
hysteresis is observed.  On the other hand if the experiment is performed after the 
concentration of surfactants has been allowed to build up, $\Delta S$ can become as large 
as 80 $\mu$m.  Note that even this large amount of hysteresis is orders of magnitude 
smaller than the typical value of the straw height and therefore does not significantly 
affect the $S_u$ vs. $Q$ curves for the equilibrated system.  

In conclusion, it is observed that allowing the surfactant concentration to 
equilibrate over a period of days has a relatively small effect (20$\%$ increase) on the $S_u$ measurements 
(figure \ref{fig:Su_vs_Q_clean/dirty}) and, when the propper precautions are taken,
no effect on the scaling relations (figure \ref{fig:scaling_clean/dirty}).

\end{document}